\documentclass[12pt,a4paper]{elsarticle}

\usepackage{lineno,hyperref}
\modulolinenumbers[5]

\journal{International Journal of Heat and Mass Transfer}

\usepackage{amsmath,amssymb,amsthm,amsfonts}
\usepackage{siunitx}	
\usepackage{subcaption}
\usepackage{float}
\usepackage{verbatim}
\usepackage[font=small]{caption}
\usepackage{tikz}
\usetikzlibrary{patterns}
\usepackage{multirow}
\usepackage{amsmath,environ}
\usepackage{url}

\makeatletter
\NewEnviron{dblequation}
 {\expandafter\dbl@equation\BODY\@nil}
\def\dbl@equation#1&#2\@nil{%
  \begin{equation}
  \makebox[\dimexpr\displaywidth-6em]{%
    \makebox[.5\displaywidth]{$\displaystyle#1$}%
    \makebox[.5\displaywidth]{$\displaystyle#2$}%
  }
  \end{equation}%
}

\usepackage{color}

\def\p{\partial}

\def\({\text{\huge (}}
\def\){\text{\huge )}}

\def\]{\text{\huge ]}}
\def\[{\text{\huge [}}

\def\dd{\textrm{d}}

\newcommand{\bi}{\begin{itemize}}
\newcommand{\ei}{\end{itemize}}
\newcommand{\be}{\begin{equation}}
\newcommand{\ee}{\end{equation}}
\newcommand{\ba}{\begin{align}}
\newcommand{\ea}{\end{align}}

\newcommand\nc{\newcommand}
\nc\pad[2]{\frac{\p #1}{\p #2}} \nc\padd[2]{\frac{\p^2 #1}{\p
{#2}^2}} \nc\nd[2]{\frac{\textrm{d} #1}{\textrm{d} #2}} \nc\pat[2]{\frac{D #1}{D
#2}} \nc\ov{\overline} \nc\degree{^{\circ}} \nc\ord[1]{{\cal
O}(#1)} \nc\ra{\rightarrow} \nc\Ra{\Rightarrow} \nc\dint{{\mbox ~
d}}

\newcommand{\bea}{\begin{eqnarray}}
\newcommand{\eea}{\end{eqnarray}}
\newcommand{\beas}{\begin{eqnarray*}}
\newcommand{\eeas}{\end{eqnarray*}}
\newcommand{\Pe}{\text{Pe}}
\newcommand{\Da}{\text{Da}}
\newcommand{\units}[1]{$^\mathrm{#1}$}

\begin{document}

\begin{abstract}
A mathematical model describing the erosion or leaching of a solid material by a flowing fluid in a column is developed. This involves an advection-diffusion equation coupled to a linear kinetic reaction describing the mass transfer between the solid and fluid. Two specific cases are analysed, the first where the extracted material has the same saturation solubility and  rate of mass transfer throughout the process, the second where the solubility switches after a certain amount of erosion. In the first case there are only two model unknowns, the solubility and mass transfer coefficient, in the second there is a third unknown, the second solubility.
Exploiting the fact that erosion is a slow process (relative to the flow rate) a perturbation solution based on the smallness of the amount removed is developed to describe the concentration and radius throughout the column. From this an analytical expression for the extracted fraction is obtained. The extracted fraction has a large linear section which results in a simple calculation to estimate the initial solubility from a very few or even a single data point. The remaining unknowns may also be easily calculated from the formula and later data points. A numerical solution, using finite differences, is developed to verify the perturbation solution. The analytical solution is also verified against experimental data for the removal of lanolin  from wool fibres with a supercritical CO$_\text{2}$/ethanol solvent. Values for the mass transfer rate and two solubilities are obtained for different pressures and shown to provide excellent agreement with a series of experimental results for the extracted fraction. 
\end{abstract}


\begin{frontmatter}
\title{Modelling mass transfer from a packed bed by fluid extraction}

\author[1]{Timothy G. Myers\fnref{myfootnote}}
\author[2]{Abel Valverde}
\author[3]{Maria Aguareles}
\author[4]{Marc Calvo-Schwarzwalder}
\author[1,5]{Francesc Font}

\address[1]{Centre de Recerca Matem\`atica, Campus de Bellaterra, Edifici C, 08193 Bellaterra, Barcelona, Spain}
\address[2]{Department of Chemical Engineering, ETSEIB, UPC, Diagonal 647, 08028, Barcelona, Spain}
\address[3]{Department of Computer Science, Applied Mathematics and Statistics, Universitat de Girona, Campus de Montilivi, 17071 Girona, Catalunya, Spain}
\address[4]{College of Natural and Health Sciences, Zayed University, PO Box 144534 Abu Dhabi, United Arab Emirates}
\address[5]{Department of Fluid Mechanics, Universitat Polit\`ecnica de Catalunya - BarcelonaTech, Barcelona 08019, Spain}

\fntext[myfootnote]{Corresponding author: tmyers@crm.cat}
\date{\today}

\end{frontmatter}

\section{Introduction}

The extraction of a material which is somehow attached to or embedded in a solid by the action of a flowing fluid   has countless applications in both nature and industry. Erosive processes occur naturally whilst many essential oils are produced through desorption  caused by the injection of solvent. In manufacturing processes solvent driven  extraction is a widely used technique. In the review of \cite{Huang12} a variety of examples of solvent extraction are described, including from leaves, seeds, roots, mushrooms and even cow brains. Essential oils, produced via extraction techniques,  are used in  areas such as food, pharmaceutical, cosmetic or perfume industries \cite{Bakkali2008}, natural dyes in the textile industry \cite{Sanda2021} or natural components for medical applications \cite{Zhang2018}. Lanolin, which coats sheep wool, has applications in pharmaceuticals, cosmetics, coatings, rust-proofing and moustache wax \cite{Thewlis1977,WikiLanolin}. Alongside the range of extracts there is also a variety of solvents. In recent years
supercritical fluids have come to the fore for reasons of cost, non-flammability, toxicity or availability, especially when compared to other commonly used petroleum-based solvents such as hexane or benzene \cite{Zhang2018,Sovova94,McHugh13,DeSimone03,Cheng11,Rajaei05,Xu00}.

The basic mathematical framework for extraction or erosion consists of a mass balance for the extract within the solvent coupled to a transfer model between the solid and solvent. In the context of column extraction various forms of this framework may be found in the literature, from coupling two differential equations \cite{Veress94}, coupled mass balances in both the solid and fluid phases \cite{Sovova94,Reverchon99,Perrut97,Sovova05,Roy96} to accounting for a moving solid-fluid interface representing the desorption of the material into the solvent \cite{ValverdeReca20,Valv19,Park75,Goto96,Levenspiel99,Fior09,Rai14}. 
Solutions are invariably numerical, via some form of  discretization \cite{Reverchon99,Perrut97,Goto96,Fior09,ValverdeReca20} 
or a polynomial approximation \cite{Valv19}. However, some simple analytical approximations have been presented which involve the neglect of spatial variation  \cite{Sovova94,Rai14} (so  variables depend only on  time). 
An extensive review of  models  can be found in \cite{Huang12}. 

Many mathematical models analogous to those describing extraction may be found in the literature of deposition processes,  for example, in
the removal from fluids containing  emerging contaminants, volatile organic compounds, CO$_\text{2}$, dyes and salts \cite{Patel19,Ahmed18}. While the chemistry may be radically different to extraction the difference in mathematical models may be as simple as a change of sign in the mass source/sink term. Consequently, there is the possibility of knowledge transfer from this field while the results of the present study may be adapted to aid in the understanding of contaminant removal.

As is the nature of models all involve degrees of approximation and assumptions,
such as averaging over the column cross-section, neglect of thermal effects or assuming negligible amounts of extract so that the solvent properties and flow are unaffected.
Certain assumptions are based on solid physical grounds while others are motivated by a desire to obtain a tractable model.  In a study of contaminant removal by adsorption \cite{Myer20a,Myer20b} rather than starting from accepted models the authors started from the basic conservation laws, rigorously identifying negligible terms and justifying all assumptions made, with the result that a number of errors in accepted models were identified. It was discussed how these errors  can lead
to problems in determining system parameters and, importantly, in the scaling-up of experiments. The corrected model of
\cite{Myer20a,Myer20b} removed many of the issues caused by these errors.

In the current study we will take the same approach as
\cite{Myer20a,Myer20b}, starting from conservation of mass and then non-dimensionalising to identify dominant and negligible terms. To keep the model general certain standard assumptions are avoided. One such assumption is that the amount extracted is small compared to the total solid mass: neglecting this assumption leads to a variation of particle radius, void fraction and velocity along the column. For cases where the amount extracted is indeed small compared to the solid the model reduces to a more standard form. During the development of the mathematical model we make no assumption on the type of solvent or solid material. However in the results section we will verify the model against experimental data for the removal of lanolin from wool via a supercritical CO$_\text{2}$-ethanol fluid.

\section{Derivation of governing equations}\label{GovSec}

We begin by stating a number of assumptions made in the development of the present model:
\begin{enumerate}
    \item The concentration/density of material attached to the core remains constant throughout the process.

    \item This material is attached to the outside of the core rather than being contained within a porous media. In this way the reduction of solid area is directly related to mass removal.

    \item The mass removed is much less than the mass of liquid passed through the column (but may be of a similar order to the amount of solid available for extraction).

    \item The inlet mass flux is  constant.
\end{enumerate}

The first assumption is consistent with a process where the solvent does not diffuse into the material to be removed, so there exists a sharp boundary between the two. This approach is employed in the shrinking core model of \cite{Goto96}.
The second assumption is valid for the processes of interest in this study, where the material to be extracted surrounds a solid core. Of course there are situations where the material is also inside a solid, porous core, for example in the Broken Intact Cell (BIC) model of \cite{Sovova94}. The present model could  be adapted to this situation, where the mass transfer rate  depends on where material is extracted. In \S \ref{LanSec} we deal with a two solubility model, where the radius changes with time for both solubilities. A BIC type model is a simplified version of this, where the radius is fixed for the second solubility phase. 
The third  assumption is consistent with a standard extraction, erosion or leaching processes. All are slow processes so that the concentration of removed material within the carrier fluid must always be small. Taking the example of \cite{Valv19,Eychenne01}  a typical result involves the removal of 3g lanolin with more than 3kg solvent. Since the volume fraction of eroded material is of the order one thousandth that of the solvent it will have a negligible effect on 
solvent properties, such as density and viscosity. The final assumption is related to the experimental setup, whereby a flow meter controls the mass flux. An alternative approach could be to use a fixed pressure drop system, where the flux then varies as material is eroded. We will not investigate this latter situation.

 \begin{table}[h!]
     \caption{Notation and key subscripts used in this work.}
     \label{tab:nomenclature}
     \begin{tabular}{cl}
      Parameters\\ 
      \\
      $A^*$ & Cross-sectional area (m\units{2})\\
      $M^*$ & Mass per unit length (kg/m)\\
      $M^*_{tot}$ & Total mass available for extraction (kg)\\
      $c^*$ & Concentration density of eroded material (kg/m\units{3})\\
      $x^*$ & Distance from the column inlet (m)\\
      $L^*$ & Length of the bed (m)\\
      $t^*$ &  Time (s)\\
      $u^*$ & Flow velocity (m/s)\\
      $R^*$ & Average radius of the core plus coating  (m)\\
      $\rho^*$ & Density (kg/m\units{3})\\
      $\epsilon$ & Void fraction  \\
      $D^*$ & Mass diffusivity (m\units{2}/s)\\
      $k^*$ & Kinetic or mass transfer coefficient (m/s)\\
      $\dot{m}^*$ & Mass flux of solvent (kg/s)\\
      $X^*$ & Extracted fraction  \\ \\
      Subscripts \\ 
      \\
      \end{tabular}
       \begin{tabular}{clccl }
      $b$ & Bed or column &\hspace{0.5cm}  & $i$ & Initial value  \\
      $c$ & Core material & & $s$ & Solvent\\
      $e$ & Material available for extraction & & $v$ & Void\\
       \\
      \hline
    \end{tabular}
\end{table}

\subsection{Derivation of the general equations}
\begin{figure}
  \centering
  \includegraphics[angle=0,width=0.45\textwidth]{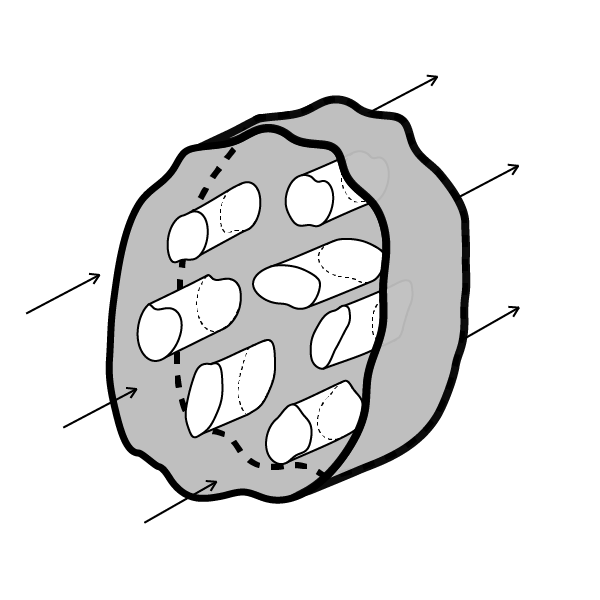}  \includegraphics[angle=0,width=0.45\textwidth]{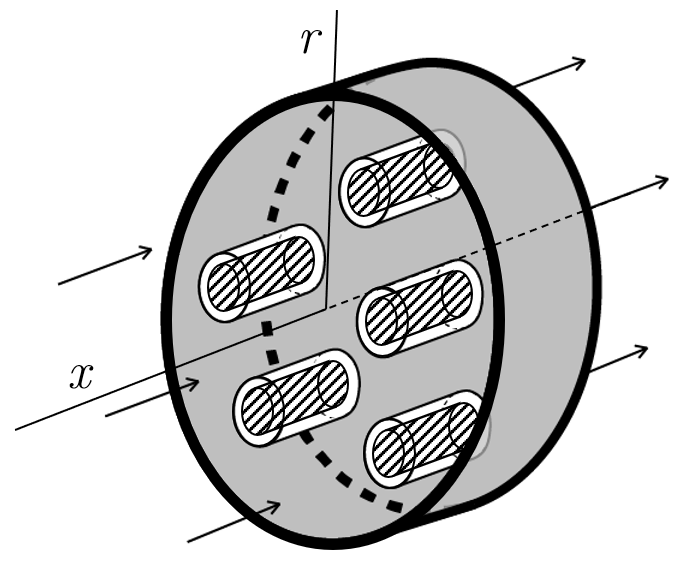}
  \caption{On the left, the diagram showing a cross-section of an arbitrary profile experimental setup. On the right, the circular cross-section column and solids  considered in \S \ref{CircSec}. The total bed circular area is A$^*_b$, the grey area inside the bed is the void area, A$^*_v$, the white area in the cylinders is the area available for extraction, A$^*_e$, and the striped area is the core, A$^*_c$.}
  \label{fig:RandShape}
\end{figure}

To specify the model we consider a packed bed with a constant inner profile. It is filled with a coated material where the inner part of the material, denoted the core, maintains a constant volume while the coating is slowly removed. The total internal cross-sectional  area is $A_b^*$, where $^*$ indicates a dimensional quantity. The solid material is our main object of interest It comprises the core and its coating and occupies  a cross-sectional area $A^*$ which is split into a constant area occupied by the core $A_c^*$ and a variable area $A_e^*$ consisting of the material to be removed; hence $A_e^*+A_c^*=A^*$. The void area $A_v^*$ is such that $A^* + A_v^* = A_b^*$. 
The configuration is depicted in Figure \ref{fig:RandShape}a. The notation used in the paper as well as the subscripts to identify where the values are defined are given in Table \ref{tab:nomenclature}.

The process is such that solvent entering at the inlet, $x^*=0$, is free of the material to be removed.
Moving along the bed the solvent slowly picks up small amounts of material until it either becomes saturated or exits the bed.
Due to the mixing caused by the presence of numerous, randomly packed solids, radial variation is small so that properties such as the concentration may be taken to vary purely with distance from the inlet and time, $c^*=c^*(x^*,t^*)$.
This approach is analogous to applying an averaging process over the cross-section. Since the packing is random the cross-sectional area will vary along the column, even at $t^*=0$, consequently we must think of the variables as a form of ensemble average, that is, an average taken over various cross-sections.
The extraction from the solid has consequences for a number of quantities. Erosion corresponds to mass loss from the solid and so $A_e^*=A_e^*(x^*,t^*)$. Since the core and bed areas are fixed $A^*(x^*,t^*)=A_c^* - A_e^*(x^*,t^*)$, $A_v^*(x^*,t^*)= A_b^*-A^*(x^*,t^*)$ are variable. Then the void fraction $\epsilon(x^*,t^*)=A_v^*(x^*,t^*)/A_b^*$ is also a variable quantity. Finally, with a varying void region conservation of mass  dictates that the interstitial velocity $u^*=u^*(x^*,t^*)$.

A straight-forward mass balance for the eroded material in the solvent over any given cross-section leads to
\bea
\pad{M^*}{t^*}     + \pad{}{x^*} (u^* M^*) = \pad{}{x^*} \left( D^* \pad{M^*}{x^*}  \right)  -  \pad{M_e^*}{t^*} \, .
\label{ceMB1}
\eea
Equation \eqref{ceMB1} states that the mass of eroded material within the solvent varies due to advection, diffusion and the rate at which eroded material enters the fluid. The mass per unit length, $M^*$, may be defined in terms of the concentration, $c^*$, of eroded material in the solvent $M^* = A_v^* c^* = \epsilon A_b^* c^*$. The diffusion coefficient depends on the void fraction and so is also variable.

The total mass of material per unit length available for extraction  is denoted $M_e^*$ where $\partial M_e^*/\partial t^* \le 0$. If $\rho_e^*$ is the constant density of this  material when attached to the core then
\bea
M_e^*=\rho_e^* A_e^* \quad \Ra \quad \pad{M_e^*}{t^*}=\rho_e^* \pad{A_e^*}{t^*} \, .
\eea

The extraction process may also be considered in terms of a kinetic reaction. Assuming a linear form we may write
\bea \label{linkineq}
\pad{M_e^*}{t^*}=- k^*  (c_s^*-c^*) \delta A_e^* \, ,
\eea
where $\delta A_e^*$ represents the total boundary of the solid area available for the reaction (or in this case mass transfer) and $c_s^*$ is the saturation concentration. The kinetic coefficient $k^*$ provides information about the extraction rate. This will vary with the ambient conditions. Both $k^*, c^*_s$ are usually unknown.

As stated in \S \ref{GovSec} a lot of solvent is required to remove small amounts of material, so the solvent mass flux is much greater than the flux of eroded material. Consequently we may state that the interstitial velocity
\bea
u^*(x^*,t^*) = \frac{\dot{m}^*}{\rho_s^* \epsilon(x^*,t^*) A_b^*} \, ,
\eea
where $\dot{m}^*$ is the mass flux of solvent and $\rho_s^*$ the solvent density.
For future reference we observe that the product $\epsilon u^*$ is constant provided $\dot{m}^*$ is constant. If $\dot{m}^*$ varies with time then the product $\epsilon u^*$ is also a function of time.

In order to proceed we must define the typical shape of the bed and  solid. The reason for this being that the mass balance is in terms of an area while the kinetic reaction depends on the surface exposed to solvent, which is the boundary of the solid material. The relation between the solid area and its boundary depends on the shape.  In Figure \ref{fig:RandShape}a) we depict  a deliberately random shape. Bed shapes can take a variety of forms while extraction from leaves, flowers, peel etc will involve many forms of solid cross-section. Fibres aligned along the bed axis will have a circular cross-section but if they are not aligned with the bed walls or are subject to bending the cross-section will be non-circular.

A common simplification is to think in terms of equivalent solid materials with a circular cross-section and then to determine an average radius depending on the void fraction. We will follow this approach in the following section and further assume a circular cross-section bed.

\subsection{Circular cross-section bed and solid}\label{CircSec}

From now on we will focus on a circular cross-section bed, with internal radius $R_b^*$ since this is the most common shape studied in the literature. In terms of the solid area it is standard to work in terms of an average radius, circular cross-section material, as depicted in Figure \ref{fig:RandShape}b). If a typical column cross-section contains $n$ solid components  which occupy a fraction $(1-\epsilon)$  then the average radius of the core material plus the coating, $R^*$, is defined through
\bea
A^*= (1-\epsilon) \pi R_b^{*2} = n \pi R^{*2} \, .
\eea
The radius of the core is denoted $R_c^*$.

Since $n$ is constant we may rearrange the above expression to define it in terms of the initial values  and also to relate the void fraction $\epsilon(x^*,t^*)$ to the bed and solid areas
\bea
\label{nepsdef}
n &=& \frac{(1-\epsilon(x^*,t^*))R_b^{*2}}{R^{*2}(x^*,t^*)} = \frac{(1-\epsilon_i)R_b^{*2}}{R^{*2}_i}
\, , \\
\epsilon(x^*,t^*) &=& 1-  n \frac{   R^{*2}(x^*,t^*)}{  R_b^{*2}}= 1-  (1-\epsilon_i) \frac{   R^{*2}(x^*,t^*)}{  R_i^{*2}}
 \, ,\label{epseq}
 \eea
where subscript $i$ indicates the initial value and it is assumed that $R^*(x^*,0) = R_i^*$, $\epsilon(x^*,0)=\epsilon_i$ are constant throughout the column.
The mass available for extraction may now be written
\bea
M_{e}^*(x^*,t^*)= \rho_e^* A_e^* = \rho_{e}^* n \pi (R^{*2} -R_c^{*2}) \, .
\eea
Since the area of the core is constant
\bea
\label{Met1}
\pad {M_e^*}{ t^*} = 2 \rho_e^* n \pi R^* \pad{R^*}{t^*}  \, ,
\eea
determines the source term. For extraction the radius decreases with time, hence $\partial R^*/\partial t^* < 0$.

The kinetic reaction involves the boundary of the solid material and so from \eqref{linkineq} we may write
\bea
\label{Met2}
\pad {M_e^*}{ t^*} = - k^*  2n\pi R^*  (c_s^*-c^*) \, ,
\eea
where $2 n \pi R^* $ is the surface area (per unit length) of the solid  available for the reaction/mass transfer. This holds whenever $c^* \le c_s^*$, $R^* \ge R_c^*$ that is extraction stops if the fluid becomes saturated or when all available material has been removed from the core.
Equating the two mass loss expressions gives
\bea
\label{Reeq}
  \pad{R^*}{t^*} = - \frac{k^*}{\rho_e^*} (c_s^*-c^*) \, .
\eea

The mass per unit length of extracted material in the fluid may be written as $M^*=\epsilon \pi R_b^2 c^*$ then, after noting that $R_b^*$ is constant,  the mass balance \eqref{ceMB1} becomes
\begin{equation}
\begin{split}
\label{Mbfinal1}
 \pad{ }{t^*}  (\epsilon c^*) + \pad{}{x^*} \left(u^* \epsilon c^* \right)  & =  \pad{ }{x^*} \left(D^* \, \pad{}{x^*}(\epsilon c^*)\right) -  \frac{1}{ \pi R_b^{*2}} \pad{M_e^*}{t^*} \\
 & =  \pad{ }{x^*} \left(D^* \, \pad{}{x^*}(\epsilon c^*)\right) +  \frac{2nk^*}{R_b^{*2}} R^*(c^*_s-c^*)
  \, ,
\end{split}
\end{equation}
where the mass source is defined by either of equations (\ref{Met1},\ref{Met2}), here we have taken the latter, and $\epsilon$ satisfies \eqref{epseq}.
In comparison a typical model from the literature takes the form
\begin{equation}
\label{StandardEq}
\epsilon \pad{c^* }{t^*}    + \epsilon u^*\pad{c}{x^*}   = \epsilon D^*\padd{c^* }{x^*}+   k^* \Gamma^*  (c^*_s-c^*)
  \, ,
\end{equation}
where $\Gamma^*$ represents a constant surface area parameter. The form of equation \eqref{StandardEq} corresponds, for example, to the \lq new' model of \cite{Fior09}, the \lq general fluid phase mass balance' in the review of \cite{Rai14} (after correcting for an error in the time derivative and converting from superficial to interstitial velocity), the \lq shrinking core model' of \cite{Goto96} and many others.

Comparison of (\ref{Mbfinal1}, \ref{StandardEq}) clearly demonstrates  key differences such as
\begin{enumerate}
    \item The variation of $\epsilon, u^*, D^*$ requires these quantities to be included within the derivative terms. Their dependence on $R^*$ makes the equations nonlinear and hence significantly more complex to solve.
    \item The source term in \eqref{StandardEq} is  expressed in terms of a specified, constant surface area, from \eqref{Mbfinal1} we see it is proportional to $R^*(x^*,t^*)$ and so variable, providing another source of nonlinearity.
\end{enumerate}

To close the system requires boundary and initial conditions.
Initially no material has  been extracted and the column is filled with a fluid that does not act as a solvent.
The first condition may be stated as $R^*(x^*,0)=R_i^*$. For the second we may apply various options. First, we could treat this as a moving boundary problem where no initial condition is imposed on $c^*$ since there is no eroded material within the column. For $t>0$ we then have an interface at $x^* = u_i^* t^*$, behind this there is solvent and we neglect the region ahead of the interface. Only when solvent reaches the outlet do we apply the system throughout the column. An alternative approach, which permits an initial condition, would be to treat the liquid  within the column at $t^*=0$ as a saturated fluid since it does not result in any extraction, i.e. define $c^*(x^*,0)=c^*_s$.
This prevents the mathematical model from predicting extraction before solvent passes through the column. If we were to naively write $c^*(x^*,0)=0$, to indicate there is no solvent in the column, then at $t^*=0$ the model would indicate that liquid already in the column would immediately erode material everywhere (and at the fastest rate possible).

As a consequence of the $R^*$ condition we may define
$\epsilon(x^*,0) = \epsilon_i^*= 1-n {R_i^{*2}}/{R_b^{*2}}$.
Since the incoming solvent is free of eroded material, $c^*(0^-,t^*)=0$ continuity of flux at the inlet requires
\bea
\label{InletCond}
0= \left. \left(u^* \epsilon c^* - D^* \pad{}{x^*} (\epsilon c^*)\right) \right|_{x^*=0^+} \, .
\eea
In the limit where $D^* \ra 0$ this may be simplified to $c^*(0,t^*)=0$. This inlet condition is based on an assumption that extraction is occurring in the vicinity of the inlet. However, if the solid material is stripped, such that $R^*=R^*_c$, then
extraction only occurs for positions $x^* > s^*(t^*)$ where beyond $x^*=s^*(t^*)$ there remains material available for extraction $R^* > R^*_c$. In which case for $x^* \ge s^*(t^*)$ we apply
\bea
c^*(s^*(t^*),t^*)=0 \, , \qquad R^*(s^*(t^*),t^*)=R^*_c \, .
\eea
For $x^* \le s^*(t^*)$ functions take the constant values $c^*(x^*,t^*) = 0, R^*(x^*,t^*)=R^*_c$.

The diffusion term of \eqref{Mbfinal1} is second order thus requiring a second boundary condition on concentration. However, since advection dominates over diffusion and so information is advected in the positive $x$ direction this condition will have little effect on the results. The full condition at the outlet should involve matching the flux expression across the boundary. However at the outlet the form of the flow is not clear and will depend on the particular experimental setup. Consequently here we assume that whatever the concentration on leaving the column it remains the same just outside the exit and impose
\bea
\label{OutletBC}
 \pad{c^*}{x^*}(L^*,t^*) = 0 \,  .
\eea
Effectively this condition indicates zero diffusive flux at the outlet, which is consistent with the fact advection dominates over diffusion, this will become  apparent in the non-dimensional analysis below. The neglect of the diffusive flux then raises the question why, if diffusion is generally negligible, include it in the inlet condition? The reason being that at the start of the process the clean fluid enters at the inlet and encounters a porous matrix packed with the maximum amount of material  available for extraction. The concentration gradient will therefore take its greatest value at early times at the inlet and this is when diffusion may  play a significant role. The same cannot be said for the outlet where the fluid containing the extracted material exits the column and the extraction process ends, changes across this boundary will therefore be small. In \S \ref{aproxS} we verify this showing that the early time solution has an outlet gradient approximately $0.005$ smaller than the inlet gradient. A second verification is provided in the results section by the close agreement between the perturbation solution (which neglects diffusion and hence only employs the inlet condition) and the full numerical solution.

\subsection{Extracted fraction}

The extracted fraction refers to the mass of extracted material passing through the outlet at a given time divided by the total amount of material initially available for extraction. It is thus a dimensionless quantity varying between 0 and 1. However, as it will be conveniently scaled in \S\ref{sec:nondim} along with  other relevant quantities, we will refer to it using the $^*$ notation.
Here we define the extracted fraction as
\bea
\label{XsDef}
X^*(t^*) =\frac{\int_0^{t^*} u^* M^*(L^*,t^*)\,\dd t^*}{M_{tot}^*}=\frac{\epsilon_i  u_i^* A_{b}^* \int_0^{t^*}  c^*(L^*,t^*) \, \dd t^*}{M_{tot}^*  } \, ,
\eea
where we have used the fact $\epsilon u^*= \epsilon_i u_i^*$ is constant and the total mass available for extraction is denoted $M^*_{tot}$. If the initial distribution of mass is  constant then $M_{tot}^* = M_e^*(x^*,0) L^*$ otherwise
\bea
M^*_{tot} = \int_0^{L^*} M_e^*(x^*,0) \, \dd x^* \, .
\eea
Subject to the circular cross-section model we have $M_{tot}^*  =  \rho_e^* n \pi (R_i^{*2}-R_c^{*2}) L^*$. Equivalently we may simply write down the available mass when provided from experimental data.

An alternative would be to define the extracted fraction in terms of the mass actually extracted during the experiment in which case the denominator of \eqref{XsDef} matches the numerator but with the upper bound of the integral replaced by $t_f^*$,
where $t_{f}^*$ represents the time at which the experiment ends. This form  accounts for the material actually extracted, which may not be the same as the available material. In the following we will use \eqref{XsDef} with $M_{tot}^*  =  \rho_e^* n \pi (R_i^{*2}-R_c^{*2}) L^*$.

\section{Non-dimensional form}\label{sec:nondim}
We begin by scaling quantities with appropriate values
\bea
{x} = \frac{x^*}{{\cal L}^*} \, , \, \,\,  {t} = \frac{t^*}{\tau^*} \, ,\, \,\, {u} = \frac{u^*}{u_i^*} \, ,\, \,\,  {c}= \frac{c^*}{c_s^*} \, ,\, \,\, {R} = \frac{R^*}{R_i^*} \, ,\, \,\, X = \frac{M_{tot}^* X^*}{\epsilon_i u_i^* A_b^* c_s^* \tau^*}\, ,
\, \,\,
 D = \frac{D^*}{D_i^*} \, .
\eea
The void fraction is already non-dimensional and of order unity, so we make no scaling. The scaling of the particle radius is chosen so that the radius is now $R \in [R_c,1]$, where $R_c=R_c^*/R_i^*$ is the scaled radius of the fiber core.
The interstitial velocity scale is chosen from the initial inlet flux $u_i^* = \dot{m}^*/(\epsilon_i \rho_s^*\pi R_b^{*2})$. Although the extracted fraction is a non-dimensional quantity we impose the above scaling to simplify the form of expression.

In the new variables, the change in radius gives
\bea
\label{ReqnND}
  \pad{R}{t} = - \frac{k^*c_s^* \tau^*}{\rho_e^* R_i^*} (1-c) = -(1-c)\, ,
\eea
provided we choose $\tau^*= \rho_e^* R_i^*/(k^*c_s^*)$. This corresponds to working on the time-scale of extraction, as opposed to the flow time-scale.
The non-dimensional mass balance for the eroded material is
\bea
\label{cfinaleq_pre}
\frac{{\cal L}^*}{u_i^*\tau^*} \pad{}{t}(\epsilon c) + \pad{}{x}( u\epsilon c) &=& \frac{D_i^*}{{\cal L}^*u_i^*} \pad{ }{x}\left( D \pad{}{x}\big(\epsilon c\big)\right)  +   \frac{2 n k^*  {\cal L}^*R_i^*}{  R_b^{*2} u_i^*} R (1-c)  \, ,
\eea
where the quantities ${\cal L}^*/(u_i^*\tau^*)$ and $D_i^*/({\cal L}^*u_i^*)$ represent the ratio of the extraction rate to the advective mass flow rate and the relative importance of diffusion to advection and thus they can be respectively classified as a form of Damk\"{o}hler number, $\Da$, and the inverse  Pecl\'{e}t number $\Pe^{-1}$.
Upon identifying the length-scale over which the extraction takes place as ${\cal L}^* = u_i^*  R_b^{*2}/(2nk^* R_i^*) =   R_i^* u_i^*/(2 (1-\epsilon_i) k^*)$ and noting that for a fixed inlet flux $\epsilon u = \epsilon_i$, \eqref{cfinaleq_pre} becomes
\bea
\label{cfinaleq}
\Da \pad{}{t}(\epsilon c) + \epsilon_i \pad{c}{x}= \Pe^{-1} \pad{ }{x}\left( D \pad{}{x}\big(\epsilon c\big)\right) +  R (1-c)
~ .
\eea
We may replace $\epsilon$ via the nondimensional form of equation \eqref{epseq},
\bea
\epsilon&=& 1-(1-\epsilon_i)R^2 \, , \label{epseq_nd}
 \eea
so that equation \eqref{cfinaleq} now involves only 
the two primary unknowns, $c, R$. These  may  be determined via the two equations (\ref{ReqnND}, \ref{cfinaleq}).
They are subject to
\bea
 \left. \left( \epsilon_i c - \Pe^{-1}  D \pad{}{x} (\epsilon c)\right) \right|_{x=0} &=& 0 \, ,  \quad \pad{c}{x}(L,t) = 0 \, ,   \quad R(x,0)=1
 \, . \, \, \, \,
\eea
As discussed in the previous section the early time concentration may be treated as a moving boundary problem, neglecting $c$ altogether at $t=0$ since initially there is no solvent within the column or by setting $c(x,0)=1$ so that the fluid initially within the column causes no extraction (by imitating a saturated fluid). When  a part of the solid is stripped, for $x \ge s(t)$, we solve (\ref{ReqnND}, \ref{cfinaleq}) subject to
\bea
\left. \left( \epsilon_i c - \Pe^{-1}  D \pad{}{x} (\epsilon c)\right) \right|_{x=s(t)} &=& 0 \, , \qquad R(s(t),t)=R_c \, .
\eea
For $x \le s(t)$ functions take the constant values $c(x,t) = 0, R(x,t)=R_c$.

The extracted fraction is now
\bea
X =  \int_0^t  c(L,t) \, \dd t  \, .
\eea

As will be seen later, when we introduce physical parameters the Damk\"{o}hler number and inverse Pecl\'{e}t number are small, indicating that the dominant balance in equation \eqref{cfinaleq} is between advection and extraction. The solution will then be well aproximated using only these two terms (this is demonstrated more formally in the following section and appendix). However, this balance is inconsistent with the boundary condition at the exit $c_x(L,t)=0$ indicating the presence of a boundary layer there, a small region where diffusion plays an important role. This will be discussed when the numerical results are presented.


\section{Approximate solution method}
\label{aproxS}

Extraction is a slow process (in comparison to the flow rate) while mass transfer by diffusion is invariably small in comparison to fluid motion. Typically only a small fraction of the total solid region is removed during the process, which means that the final fibre radius is close to the initial value, $R_c^*\sim R_i^*$. Consequently we anticipate $\Da, \Pe^{-1} \ll 1-R^*_c/R_i^* \ll 1$ (see the values shown in \eqref{DaPeData} for the application considered in Section \S\ref{LanSec}).
In this section we exploit the difference in scales to derive approximate expressions for the radius of the  fibres, the concentration of the solid in the solvent, the void fraction in the column and the total extracted mass.

We shall consider two commonly encountered scenarios: one in which the saturation concentration remains the same 
throughout the process and a second in which the material to be extracted is composed of two different fractions, with two different saturation concentrations. The first is obviously the most common, and simplest to model, where a single solubility material coats the solid matrix. The second occurs, for example, with lanolin. The experimental results presented in \cite{Eychenne01} clearly demonstrate two different solubilities where the first fraction  contains the
light esters and free alcohols while the second has the
higher molecular weight products. The solubility of
the second being significantly lower than that of the first.  The Broken-Intact Cell (BIC) model of Sovov\'{a} \cite{Sovova94} deals with extraction from a milled material where cells that have been opened by the milling have easily accessible material which is removed first, subsequently there is a slower extraction of material protected by the intact cell walls. 

\begin{figure}[h]
  \centering
  \includegraphics[scale=0.5]{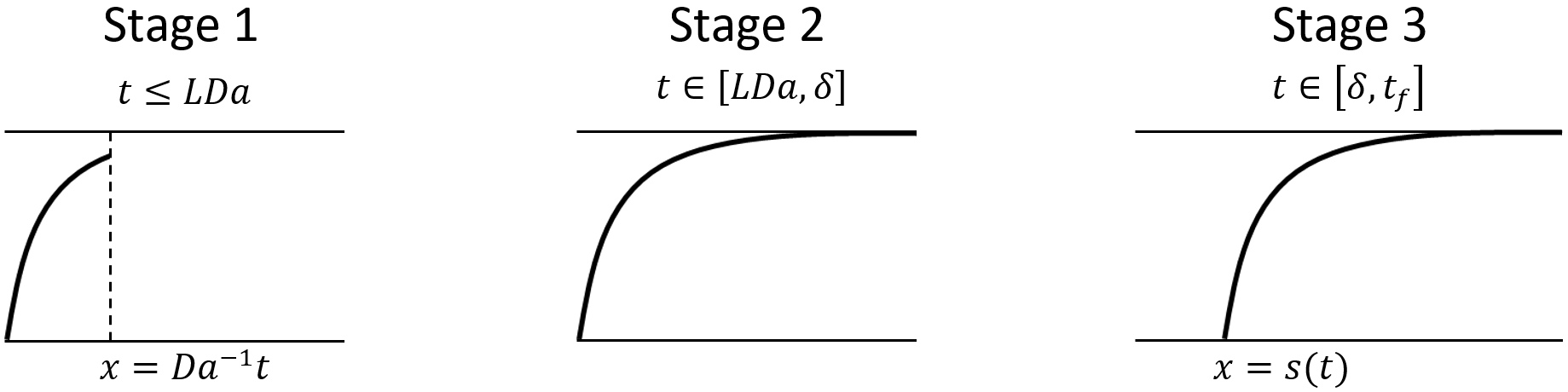}
  \caption{Representation of the evolution of the concentration during the three stages of the extraction process for a single solubility material.} 
  \label{fig:stage123Da}
\end{figure}
\subsection{Single solubility model}
 In the single solubility model the material is extracted in three stages, as depicted in Figure \ref{fig:stage123Da}.  
Approximate expressions for the concentration and radius during each stage are given below. Details of their derivation are provided in  \ref{aprox}.1 where a perturbation method is used based on the smallness of the parameter $\delta = 1-R^*_c/R^*_i=1-R_c \ll 1$. 
Taking only the leading order, {\it i.e.} neglecting $\delta$ altogether, will lead to errors of order $\delta$ while including terms of order $\delta$ leads to errors of order $\delta^2$ (in fact the application of boundary conditions, which force the perturbation to match the exact solution at the boundaries, means that errors are often much smaller than indicated by the size of neglected terms). So, for example, if $R^*_c = 0.9 R_i^*$ such that $\delta = 0.1$ then the leading order results in errors of order 10\% while the first order errors will be of order 1\%. 
\begin{description}
    \item[Stage 1.] At very small times the clean solvent starts to flow through the column but there has not been sufficient time for it to reach the outlet. An interface exists between the solvent and fluid initially occupying the column. Beyond the interface no extraction has occurred and the void fraction is $\epsilon = \epsilon_i$. This indicates that the interface moves with velocity $u=1$ (i.e. $u^*=u_i^*$). So, during Stage 1, solvent occupies the region  $x \in [0, t/\Da]$ ({\it i.e.} $x^* \in [0, u_i^* t^*]$).  Solvent reaches the end of the column at time $t=t_1=L\Da$ ($t_1^* = L^*/u_i^*$). Since extraction is slow we assume that this occurs before all soluble material has been eroded at the inlet (this will be verified later when we deal with a concrete example in Section \S\ref{LanSec}). 
    
    \item[Stage 2.] Solvent now occupies the whole column. This stage continues until all soluble material has been removed  at the inlet, that is for $t\in[L\Da, t_2]$, where $t_2 = \delta=1-R_c$ is the time required for the radius to reduce from 1 to $R_c$. Noting that $t_2= \delta \ll 1$ we observe that these early stages occupy a small amount of the total process time. 
    
  Approximate solutions for the concentration and the radius during Stages 1 and 2 are given by
    \bea
    \label{cRStage2}
    c_e(x,t)= (1-e^{-x/\epsilon_i})(1-t e^{-x/\epsilon_i}),\quad R_e(x,t) = 1-t e^{-x/\epsilon_i} \, ,
    \eea
    where $\delta$ does not appear explicitly due to a rescaling of time carried out in the appendix.
    We have denoted these solutions by $c_e, R_e$ since they represent the early time behaviour, $0 < t \le t_2$. In Stage 1, $t \in [0, t_1]$, these expressions hold over the region $x \in [0, t/\Da]$ , while for $x \ge t/\Da$ there is no solvent (so the initial values apply). 
    For Stage 2, $t \in [t_1, t_2]$,  equation \eqref{cRStage2} holds over the whole column and this lasts until  time $t_2 = \delta$ ($t_2^*= \delta \tau$).
    
    \item[Stage 3.] The inlet material has been stripped 
     and the point where $R^*=R^*_c$ now progresses through the column until all soluble material has been removed. Except very close to time $t=t_2$ the solution in this stage may be approximated by a travelling wave. Behind the wave $c=0, R=R_c$, in front the solution to first order is
    \begin{equation}
\label{cRStage3}
\begin{split}
c_l(x,t) &= \left(1- e^{-x/\epsilon_i} e^{({t}/\delta- 1)}\right) \left(1-  \delta  e^{-x/\epsilon_i} e^{({t}/\delta- 1)} \right)
\, , \\
R_l(x,t) &=   1-\delta  e^{-x/\epsilon_i} e^{({t}/\delta- 1)} \, ,
\end{split}
\end{equation}
where the subscript $l$ denotes the late time solution.
The interface position is defined by $s(t)=v(t-t_2)$, so the above solutions hold for $x \in [s(t), L]$. This stage finishes when all soluble material in the column has been extracted, that is $R(L)=R_c$, where the corresponding time may be calculated from \eqref{cRStage3}, $t_f = \delta(1+L/\epsilon_i)$ ($t_f^* = \delta(1+L/\epsilon_i)\tau$). 
\end{description}
Once  the radius is determined the void fraction $\epsilon(x,t)$ may be specified using equation \eqref{epseq_nd} while the solvent velocity $u(x,t) = \epsilon_i/\epsilon(x,t)$.

The extracted fraction is defined as the ratio of material collected at the outlet to the total amount available. Since Stage 1 accounts for the period before solvent reaches the outlet $X=0$ for $ {t} \le { {t}_1}$. Subsequently   we may write  
\begin{align}
X_e(t)  & =  \int_{ {t}_1}^{  {t}}  c_e( L,\xi) \, \dd\xi  \qquad \mbox{for}  \quad  { {t}_1} \le  {t} \le t_2\, ,  \\
X_l(t)  & =  \int_{ {t}_1}^{  t_2}  c_e( L,\xi)  \, \dd\xi + \int_{ t_2}^{ {t}}   c_l(L,\xi) \, \dd\xi  \qquad \mbox{for}  \quad  t_2 \le  {t} \le  {t}_f
\, .
\end{align}
This leads to
\begin{align}
X_e(t)  & = (1-E_L)\left( {t} - {t}_1 \right)  +\ord{\delta^2}\, , \quad \mbox{for}  \quad  { {t}_1} \le  {t} \le t_2  \label{Xe1} \\
X_l(t)  & = X_e(t_2) +
  {t} - t_2 - E_L t_2  \left(e^{({t}/t_2 - 1)}-1\right) +\ord{\delta^2} \nonumber \\ & =
  t - (1-E_L) t_1 - E_L t_2   e^{({t}/t_2 - 1)}+\ord{\delta^2} \, , \, \, \,  \mbox{for} \, \, \, t_2 \le   {t} \le  {t}_f
\, , \label{Xl1}
\end{align}
where $E_L = \exp(-L/\epsilon_i)$. Since we neglect terms of $\ord{\delta^2}$ and $t_2 = \delta > t_1$ the quadratic terms arising from integrating $t$ in the $c_e$ equation have been omitted. 
In dimensional form these become
\small
\begin{align}
\label{Xesingle}
X^*_e(t^*)  & \approx  \frac{\dot{m}^* c^*_s}{M_{tot}^* \rho_s^*} \left(1- E_L \right) \left(t^*- t_1^*  \right) \, , \quad \mbox{for}  \quad   t_1^* \le t^*\le t_2^* \, ,
\\
\label{Xlsingle}
X_l^*(t^*) &\approx \frac{\dot{m}^* c^*_s}{M_{tot}^* \rho_s^*} \left(t^* - (1-E_L) t^*_1 - E_L t^*_2   \exp\left(\frac{t^*}{t^*_2} - 1\right)\right)\, , \quad
   \mbox{for} \, \, \,   t_2^*  \le   {t^*} \le  t_2^* \left(1+\frac{L^*}{\epsilon_i {\cal L}^*}\right)
\, ,
\end{align}
\normalsize
where $E_L= \exp(- L^* /(\epsilon_i {\cal L}^*))$, ${\cal L}^* = R_i^* u_i^*/(2(1-\epsilon_i) k^*_i)$, $t_1^*= L^*/u_i^*$, $t_2^* = (R_i^*-R^*_c) \rho_e^* /(k^* c_s^*)$, $\tau^* = \rho_e^* R_i^*/(k^* c_s^*)$ and we have used the definition $u_i^* = \dot{m}^*/(\epsilon_i \rho_s^* A_b^*)$.

In practical situations  $E_L \ll 1$ (in the examples of the following sections the typical value is $10^{-3}$) in which case for the majority of the process the extracted fraction rate $d X^*/dt^* \approx \dot{m} c^*_s/(M_{tot}^* \rho_s^*)$ is constant. Hence the extraction  rate increases with an increase in the inlet mass flux and the fluid's saturation concentration or a decrease in the solvent density. These are the key factors affecting the rate, at least during the linear period (which lasts for a significant proportion of the process).

The above analytical solutions permit the verification of the outlet boundary condition. From \eqref{cRStage2} the early time concentration gradient at the inlet and outlet takes the form
\bea
\pad{c_0}{x}(0,t) = \frac{c_s^*}{\epsilon{\cal L} }(1-t^*/\tau^*) \, , \qquad
\pad{c_0}{x}(L,t) = \frac{c_s^*E_L}{\epsilon{\cal L}}(1+t^*/\tau^*-2t^* E_L/\tau^*) \, .
\eea
That is $c_{0x^*}^*(L^*,t^*) = \ord{E_L c_{0x^*}^*(0,t^*)}$ which verifies the discussion behind the neglect of the diffusive term in the flux condition of \eqref{OutletBC}.

The above solution involves the two unknowns $c_s^*, k^*$. Practically these may be determined by taking two experimental data points for the extracted fraction and then solving the two simultaneous equations numerically. Of course more data points could be used and an average taken. Using the above approximation to determine the unknown parameters is considerably simpler than solving the PDE system and then optimising the unknowns.  The only contentious issue is that, without knowing the value of $t_2^*$ (which depends on the unknowns) it is hard to say whether the data points fall into early or late time solutions. Consequently we would solve assuming the results lie in one region and then adjust if the points used were found to be outside of the resultant $t_2^*= \delta \tau$. The determination of parameter values is made significantly simpler through the observation that the extracted fraction has a linear form and hence constant slope for much of the process thus permitting $c_s^*$ to be calculated from the slope. The value of $k^*$ requires information from the nonlinear part of the process, that is for large times. 

\subsection{Model for two distinct solubilities}\label{TwoSolSec}
We now consider the situation where the solubility changes at a known value of the radius, denoted $R_w$. For $R \in [R_w, 1]$ the solubility is $c_s$, for $R \in [R_c, R_w]$ the solubility is denoted $c_w$. The early time behaviour is identical to Stages 1 and 2 of the single solubility model, but now Stage 2 ends when the inlet radius reaches $R_w$. Subsequently the region from the inlet to the point where $R=R_w$ has solubility $c_w$ while the region ahead of this, with $R>R_w$, has  solubility $c_s$. 

There are various possible solution forms at this stage, depending on the solubility and relative thickness of the two layers. Below we will detail the case where the interface between the two solubility regions reaches the outlet before the inlet region reaches $R=R_c$. This is motivated by the example studied in \S \ref{LanSec}, where  the second solubility is much lower than the first and so we anticipate all material with $c_s$ being removed before the $c_w$ section is stripped at the inlet. An alternative scenario is that the inlet is stripped before the interface between the two solubilities reaches the outlet, then the model will require two  moving fronts. The BIC model \cite{Sovova94} would require $R$ to decrease for the first stage but stay constant for the second (since material is being removed from inside intact cells). It is also possible that with the change in solubility there is a change in reaction rate, manifested through $k$. In all cases a similar analysis to that discussed below would provide the appropriate solution forms. 

Now we summarise the results of  \ref{aprox2}. 
\begin{description}
    \item[Stages 1 and 2.] In these first stages all the material extracted has the solubility $c_s$. Therefore, the concentration and radius are identical to those of Stages 1 and 2 for the single solubility model, equations \eqref{cRStage2}. In this case they hold until the solubility first changes, when $R(0, {t}_2) = R_w$. From equation \eqref{cRStage2} we determine the time when this stage ends as
\bea
\label{t2neweq}
 {t}_2 =  1-R_w  \, .
\eea
Subsequently we again seek a travelling wave solution.

\item[Stage 3.]  Now there is material to extract both before and after the front. To first order the concentration is
\bea
\label{cStage3}
 \hspace{-1.0cm} c_l(x,t)=\left\{\begin{array}{ll}
   c_w & \textrm{if $x\le s_1(t)$}  \\
    1-(1-c_w)\left(1+(1-R_w)\left(1-\exp\left(-\frac{x-s_1(t)}{\epsilon_i}\right)\right)\right)\exp\left(-\frac{x-s_1(t)}{\epsilon_i}\right) & \textrm{if $x\ge s_1(t)$}
\end{array}\right.\, 
\eea
For $x \le s_1(t)$ the solution indicates that the concentration rapidly reaches its new,  lower, saturation value $c_w$ (this could be improved by seeking higher order terms or treating the boundary layer at the inlet). The error is greatest when $s_1 \approx 0$, in the Appendix it is explained that the error decreases exponentially away from the inlet. Using the values of \S \ref{LanSec} shows a maximum error of around 10\% which will be negligible by the time the front reaches the outlet.

The radius is given by
\bea
\label{Rad1stFrontA}
R_l(x,t) = \left\{\begin{array}{ll}
R_w & \textrm{if $x\le s_1(t)$}\\
    1 - (1-R_w) e^{-(x-s_1(t))/\epsilon_i} & \textrm{if $x\ge s_1(t)$}
\end{array}\right.\, ,
\eea
and the front position
\bea
s_1(t) = \epsilon_i(1-c_w)\, \frac{t-t_2}{1-R_w}=\epsilon_i(1-c_w)\left(\frac{t}{1-R_w}-1\right)\, .
\eea
The solution $R_l = R_w$ for $x \le s_1$ is consistent with the concentration and subject to the same restrictions. 

This solution holds until $t=t_3$ such that $s_1(t_3)=L$ where
\bea
t_3 = (1-R_w) \left(1+\frac{L}{\epsilon_i (1-c_w)}\right) \, .
\eea
As a check on the solutions we note that when $R_w=R_c$, $c_w=0$ and the single solubility results are retrieved.

    \item[Stage 4.] In this stage all $c_s$ material has been removed and so again we have a single solubility problem. This is easily solved to determine
    \bea
c_{f1}(x, {t})\sim c_w(1-e^{-x/\epsilon_i})\, , \quad R_{f1}(x, {t})\sim  R_w-c_w ( {t}- {t}_3)e^{-x/\epsilon_i}\, .
\eea
Stage 4 ends when $R(0,t)=R_c$ at time
  \bea
{t}_4={t}_3+\frac{R_w-R_c}{c_w }=(1-R_w) \left(1+\frac{L}{\epsilon_i (1-c_w)}\right)+\frac{R_w-R_c}{c_w}\, .
\eea

 \item[Stage 5.] The final stage is simply a single solubility travelling wave and follows the method of Stage 3 from the previous section. Then, behind the wave $c=0, R=R_c$, and in front the concentration and radius for $ {t}_4\leq  {t}\leq {t}_f$ are now given by
\bea
c_{f2}(x, {t})&\sim & c_w(1-e^{-(x-s_2( {t}))/\epsilon_i})\, ,\\
R_{f2}(x, {t})&\sim & R_c+(R_w-R_c)(1-e^{-(x-s_2( {t}))/\epsilon_i})\, , 
\eea
where the front position is
\bea
s_2(t) = \frac{\epsilon_i c_w}{R_w - R_c} (t-t_4) \, .
\eea
The process ends when $s_2(t_f) = L$, that is
\bea
t_f =  t_4 + \frac{(R_w - R_c)L}{\epsilon_i c_w} \, .
\eea
\end{description}

Using the above expressions one can now obtain the extracted fraction for the two solubility model. As before we will neglect all terms of $\ord{\delta^2}$.
The previous single solubility result holds for $t_1 \le t \le t_2$ where $t_1=L/u$, $t_2 = (1-R_w)$,
\bea
X_e(t) =\int_{t_1}^{t} (1-E_L)(1-t E_L) \dd t = (1-E_L) (t-t_1) +\ord{\delta^2}\, .\nonumber
\eea
For $t_2<t\le t_3$ where $t_3=(1-R_w)\left(1+L/(\epsilon_i (1-c_w))\right)$,
\bea
X_l(t) &= & (1-E_L) (t_2-t_1)+ \int_{t_2}^t\left( 1-(1-c_w)E_L e^{(1-c_w)(t-t_2)/(1-R_w)}+\ord{\delta}\right) \, \dd t \nonumber \\
&= & t - t_1 (1-E_L) - (1-R_w) E_L\exp\left((1-c_w)\left(\frac{t}{1-R_w}-1\right)\right) +\ord{\delta^2}\nonumber
\, .
\eea
For $t_3<t\le t_4$ where $t_4=t_3+(R_w-R_c)/c_w$,
\bea
X_{f1}(t) &= & X_l(t_3)+\int_{t_3}^t c_w(1-E_L)+\ord{\delta^2}=t_3 - t_2 + (1-E_L)(c_w(t-t_3) - t_1) +\ord{\delta^2}\nonumber
\, .
\eea
Finally, for $t_4<t\le t_f$ where $t_f=t_4+L(R_w-R_c)/(c_w\epsilon_i)$,
\bea
X_{f2}(t) &= & X_{f1}(t_4) + \int_{t_4}^t c_w\left(1-E_L e^{c_w (t-t_4)/(R_w-R_c)}\right)+\ord{\delta^2} \nonumber\\
&= & t_3-t_2-(1-E_L)t_1+c_w(t-t_3)\nonumber\\
&&-E_L(R_w-R_c)\exp\left(\frac{c_w(t-t_3)}{R_w-R_c}-1\right)+\ord{\delta^2}\, .\nonumber
\eea
In dimensional form,
\small
\bea
X^*_e(t^*) & \approx & \frac{\dot{m}^* c^*_s}{M_{tot}^* \rho_s^*}   (1-E_L) \left(t^*-t_1^* \right) \, , \quad \textrm{for $t_1^*  \le   {t^*} \le  t_2^*$,}\, ,\label{Xestwosol}\\
X_l^*(t^*) & \approx & \frac{\dot{m}^* c^*_s}{M_{tot}^* \rho_s^*} \Bigg(
t^* - t^*_1 (1-E_L) \nonumber\\
&&- t^*_2 E_L \exp\left(\left(1-\frac{c_w^*}{c_s^*}\right)\left(\frac{t^*}{t^*_2}-1\right)\right)
\Bigg)\, , \quad
  \textrm{for $t_2^*  \le   {t^*} \le  t_3^*$,} \label{Xltwosol}\\
X_{f1}^*(t^*) & \approx & \frac{\dot{m}^* c^*_s}{M_{tot}^* \rho_s^*} \left(t^*_3 - t^*_2 + (1-E_L)\left(\frac{c^*_w}{c^*_s}(t^*-t_3^*) - t^*_1\right)\right)
\, ,\quad
  \textrm{for $t_3^*  \le   {t^*} \le  t_4^*$,}\\
X_{f2}^*(t^*) & \approx & \frac{\dot{m}^* c^*_s}{M_{tot}^* \rho_s^*}\left(t_3^*-t_2^*-(1-E_L)t^*_1+\frac{c^*_w}{c^*_s}(t^*-t^*_3)\right.\nonumber\\
&&\left.-\tau^* E_L\frac{R^*_w-R^*_c}{R^*_i}\exp\left(\frac{c^*_w R_i^*(t^*-t^*_3)}{c^*_s\tau^*(R^*_w-R^*_c)}-1 \right)\right)
\, ,\quad
  \textrm{for $t_4^*  \le   {t^*} \le  t_f^*$,}
\label{Xf2}
\eea
\normalsize
where $E_L= \exp(- L^* /(\epsilon_i {\cal L}))$, ${\cal L} = R_i^* u_i^*/(2(1-\epsilon_i) k^*_i)$, $t_1^*= L^*/u_i^*$, $t_2^* = \tau^*(1-R^*_w/R^*_i)$, $t_3^*=\tau^*(1-R_w^*/R_i^*)(1+L^*/(\epsilon_i{\cal L}(1-c_w^*/c_s^*)))$, $t_4^*=t_3^*+c_s^*(R_w^*-R_c^*)/(c_w^* R_i^*)$, $t_f^*=t_4^*+L^*c_s^*(R_w^*-R_c^*)/(c_w^*R_i^*{\cal L}\epsilon_i)$, $\tau^* = \rho_e^* R_i^*/(k^* c_s^*)$ and we have again used the definition $u_i^* = \dot{m}^*/(\epsilon_i \rho_s^* A_b^*)$.

Again we note the possibility of linear sections:  the first part with gradient  $\dot{m}^* c^*_s/(M_{tot}^* \rho_s^*)$ matches that of the single solubility model, the second has gradient  $\dot{m}^* c^*_w/(M_{tot}^* \rho_s^*)$. These regions join where the exponential term of  \eqref{Xltwosol} becomes non-negligible. In general the double solubility model is clearly more complex than the single version. In the above example there are five distinct stages, and this is just for the option where all $c_s$ material is removed before the $c_w$ section is removed at the inlet. The physical situation may be such that instead of a change in solubility there is a change in the mass transfer coefficient or the radius is fixed after the shift in solubilities. Our choice was based on the example used in \S \ref{LanSec}. A similar analysis should be possible for all other cases.

\section{Numerical solution}

In the previous section, we derived analytical solutions valid for small $\Da$, $\Pe^{-1}$ and $\delta$. In this section, we verify the accuracy of these solutions by comparing them to a numerical solution of the full model.  

For the single solubility system
the governing equations for the three stages of the process are the same, however the domain where they hold changes for each stage. These domains are $\Omega_1 = [0,l(t)]$, $\Omega_2=[0,L]$ and $\Omega_3=[s(t),L]$, for stages 1, 2 and 3, respectively, where $l(t)=t/\Da$ and $s(t)$ is unknown which has to be found as part of the solution. This suggests a different numerical strategy for each stage. However, as we will show, stages 2 and 3 can be dealt with together using a similar approach to that  previously employed  for adsorption processes in porous media \cite{Myer20a,Myer20b}. 

Before presenting the numerical strategy we define the new variable $g=\epsilon c$ and rewrite equations \eqref{cfinaleq} and \eqref{ReqnND} as
\begin{align}
\Da \frac{\partial g}{\partial t} + \epsilon_i \frac{\partial}{\partial x}\left[\frac{g}{1-(1-\epsilon_i)R ^2}\right] &= \Pe^{-1} \frac{\partial^2 g}{\partial x^2}+R\left[1-\frac{g}{1-(1-\epsilon_i)R^2}\right]\,,\label{eqnum1}
\\
\pad{R}{t} &= -\left[1-\frac{g}{1-(1-\epsilon_i)R^2}\right]\,,\label{eqnum2}
\end{align}
subject to the boundary conditions
\begin{align}
\epsilon_i \left[\frac{g}{1-(1-\epsilon_i)R^2}\right]\bigg\rvert_{x=0} - \Pe^{-1} \frac{\partial g}{\partial x}\bigg\rvert_{x=0} = 0\,,\qquad \frac{\partial }{\partial x}\left[\frac{g}{1-(1-\epsilon_i)R^2}\right]\bigg\rvert_{x=L} = 0\,.\label{eqnum3}
\end{align}
We do not specify the initial conditions yet since they change at each stage.

\textbf{Stage 1:} Equations \eqref{eqnum1}-\eqref{eqnum2} hold in the spatial domain $\Omega_1$  until the solvent reaches the end of the column at time $t_1=L\Da$. We use a popular strategy to tackle moving boundaries in numerical schemes that consists on mapping the variable domain into a fixed, unit domain using a Landau type transformation. In our case, the Landau transformation is $\xi=x/l(t)$, leading to the  unknown functions $\tilde{g}(\xi,t)=g(x,t)$ and $\tilde{R}(\xi,t)=R(x,t)$. In terms of the transformed variables, the governing equations \eqref{eqnum1}-\eqref{eqnum2} read 
\begin{align}
 \frac{\partial \tilde{g}}{\partial t}-\frac{\dot{l}\,\xi}{l\, \Da}\frac{\partial \tilde{g}}{\partial \xi} +  \frac{\epsilon_i}{l\, \Da}\frac{\partial}{\partial \xi}\left[\frac{\tilde{g}}{1-(1-\epsilon_i)\tilde{R}^2}\right] &= \frac{\Pe^{-1}}{\Da\, l^2} \frac{\partial^2 \tilde{g}}{\partial \xi^2}+\frac{\tilde{R}}{\Da}\left[1-\frac{\tilde{g}}{1-(1-\epsilon_i)\tilde{R}^2}\right]\,,\label{eqnumS11}
\\
\pad{\tilde{R}}{t}-\frac{\dot{l}\,\xi}{l} \pad{\tilde{R}}{\xi}&= -\left[1-\frac{\tilde{g}}{1-(1-\epsilon_i)\tilde{R}^2}\right]\,. \label{eqnumS12}
\end{align}
Note the only true unknowns in \eqref{eqnumS11}-\eqref{eqnumS12} are $\tilde{g}$ and $\tilde{R}$, since the position of the moving boundary, $l(t)= \Da^{-1} t$, and its derivative, $\dot{l}=\Da^{-1}$, are known. The boundary conditions \eqref{eqnum3} become   
\begin{align}
\epsilon_i \left[\frac{\tilde{g}}{1-(1-\epsilon_i)\tilde{R}^2}\right]\bigg\rvert_{\xi=0} - \frac{\Pe^{-1}}{l} \frac{\partial \tilde{g}}{\partial \xi}\bigg\rvert_{\xi=0} = 0\,,\qquad \frac{1}{l}\frac{\partial }{\partial \xi}\left[\frac{\tilde{g}}{1-(1-\epsilon_i)\tilde{R}^2}\right]\bigg\rvert_{\xi=1} = 0\,.\label{eqnumS13}
\end{align}
Initially $l(0)=0$, so the domain has zero thickness, and there are no initial conditions. To overcome this issue, we simply  initialise the code with $l(t_0)=l_0$, where $t_0$ is very small and $l_0=\Da^{-1}t_0$, and apply $\tilde{g}(\xi,t_0)=0$ and $\tilde{R}(\xi,t_0)=1$.

To solve \eqref{eqnumS11}-\eqref{eqnumS13}, we implement an explicit finite difference
scheme with suitable upwind discretisation for first-order spatial derivatives and central differences for the diffusion term. So, the time derivatives in the governing equations are approximated via
\begin{align}
\frac{\partial \tilde{g}}{\partial t}\approx \frac{\tilde{g}^{n+1}_i-\tilde{g}^{n}_i}{\Delta t}\,, \quad
\frac{\partial \tilde{R}}{\partial t}\approx \frac{\tilde{R}^{n+1}_i-\tilde{R}^{n}_i}{\Delta t}\,
\end{align}
and spatial derivatives via
\begin{align}
\frac{\partial \tilde{g}}{\partial \xi}\approx \frac{\tilde{g}^{n}_{i+1}-\tilde{g}^{n}_{i}}{\Delta \xi}\,,\quad 
\frac{\partial \tilde{R}}{\partial \xi}\approx \frac{\tilde{R}^{n}_{i+1}-\tilde{R}^{n}_{i}}{\Delta \xi}\,,\quad
\frac{\partial f}{\partial \xi}\approx \frac{f^{n}_{i}-f^{n}_{i-1}}{\Delta \xi}\,,
\quad
\frac{\partial^2 \tilde{g}}{\partial \xi^2}\approx \frac{\tilde{g}^{n}_{i+1}-2\tilde{g}^{n}_{i}+\tilde{g}^{n}_{i-1}}{\Delta \xi^2}\,,
\end{align}
where $f=g/(1-(1-\epsilon_i)R^2)$. We  ensure that all the stability conditions are satisfied.

\textbf{Stages 2 and 3:} For a compact formulation that allows  Stages 2 and 3 to be tackled together, we rewrite \eqref{eqnum1}-\eqref{eqnum2} using the the Heaviside function $H(R-R_c)$ as
\begin{align}
\Da \frac{\partial g}{\partial t} + \epsilon_i \frac{\partial}{\partial x}\left[\frac{g}{1-(1-\epsilon_i)R ^2}\right] &= \Pe^{-1} \frac{\partial^2 g}{\partial x^2}+R\left[1-\frac{g}{1-(1-\epsilon_i)R^2}\right] H(R-R_c)\,,\label{eqnumS231}
\\
\pad{R}{t} &= -\left[1-\frac{g}{1-(1-\epsilon_i)R^2}\right]H(R-R_c)\,,\label{eqnumS232}
\end{align}
where $H(R-R_c)$ disables the source term in \eqref{eqnumS231} and  \eqref{eqnumS232} altogether wherever $R-R_c=0$. These equations are subject to the boundary conditions \eqref{eqnum3}. The initial conditions are 
\begin{align}\label{eqnumS233}
R(x,t_{1})=R_{s1}(x)\,,\qquad g(x,t_{1})=g_{s1}(x)\,,
\end{align}
where $R_{s1}(x)$ and $g_{s1}(x)$ are the profiles at the end of Stage 1, $t=t_{1}$. An explicit scheme for \eqref{eqnumS231}-\eqref{eqnumS233} is used in an analogous way to Stage 1, i.e., using an upwind discretisation for the advection term, central differences for the diffusion term, and ensuring that all stability conditions are satisfied. 

\subsection{Extension to two solubilities}

The switch between solubilities does not involve a particular challenge for the numerical scheme. Using the function $\chi=\chi(R)$ that switches from $1$ to $c_{\omega}$ when $R=R_{\omega}$ (see \ref{aprox}) the system \eqref{eqnumS231}-\eqref{eqnumS232} becomes 
\begin{align}
\Da \frac{\partial g}{\partial t} + \epsilon_i \frac{\partial}{\partial x}\left[\frac{g}{1-(1-\epsilon_i)R ^2}\right] &= \Pe^{-1} \frac{\partial^2 g}{\partial x^2}+R\left[\chi-\frac{g}{1-(1-\epsilon_i)R^2}\right] H(R-R_c)\, ,\label{eqnumS231_vis}
\\
\pad{R}{t} &= -\left[\chi-\frac{g}{1-(1-\epsilon_i)R^2}\right]H(R-R_c)\,,\label{eqnumS232_vis}
\end{align}
and we may proceed as before.

\subsection{Comparison between numerics and perturbation}

\begin{figure}[htb]
\centering
\includegraphics[angle=0,width=0.5\textwidth]{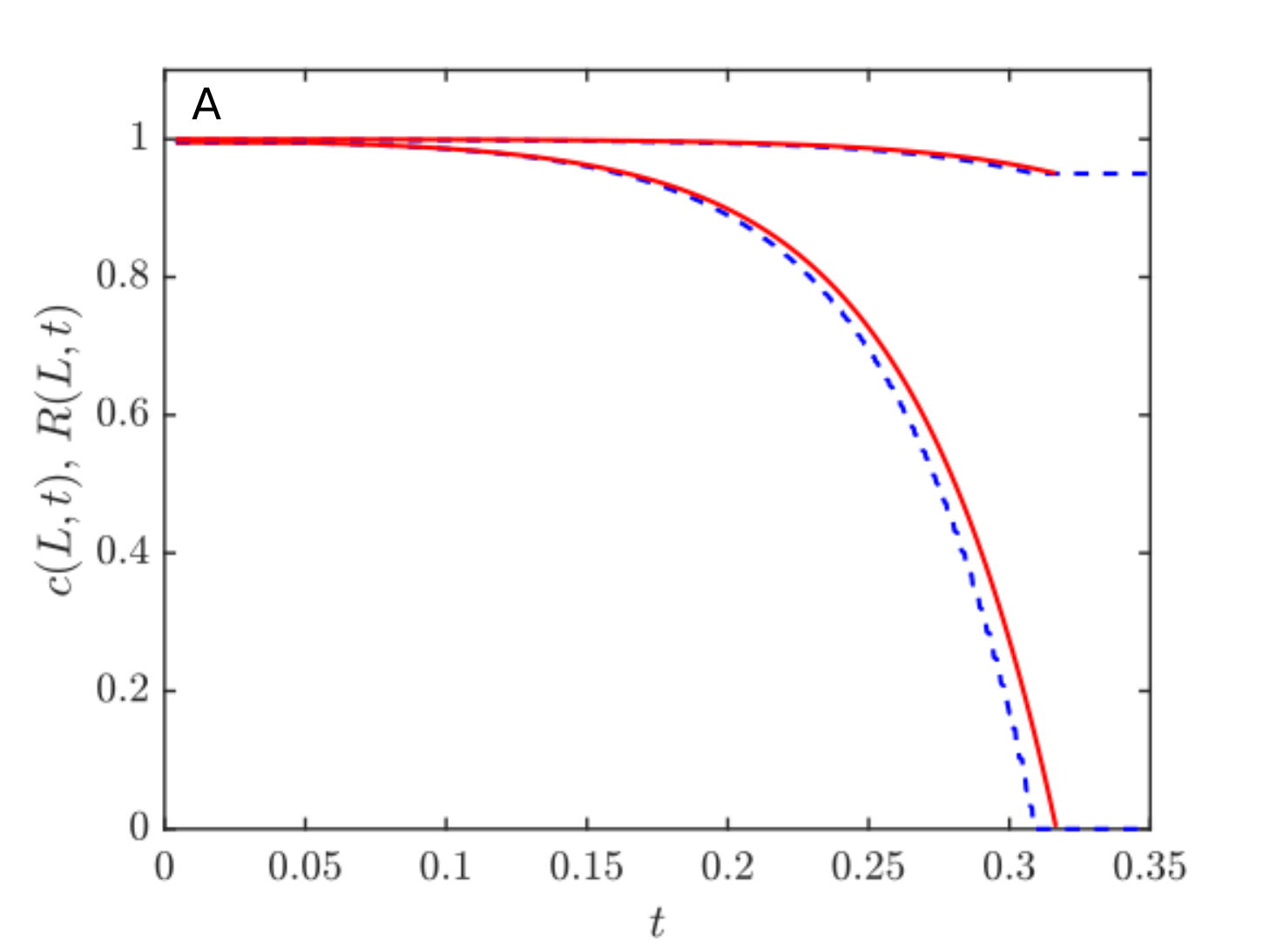}\includegraphics[angle=0,width=0.5\textwidth]{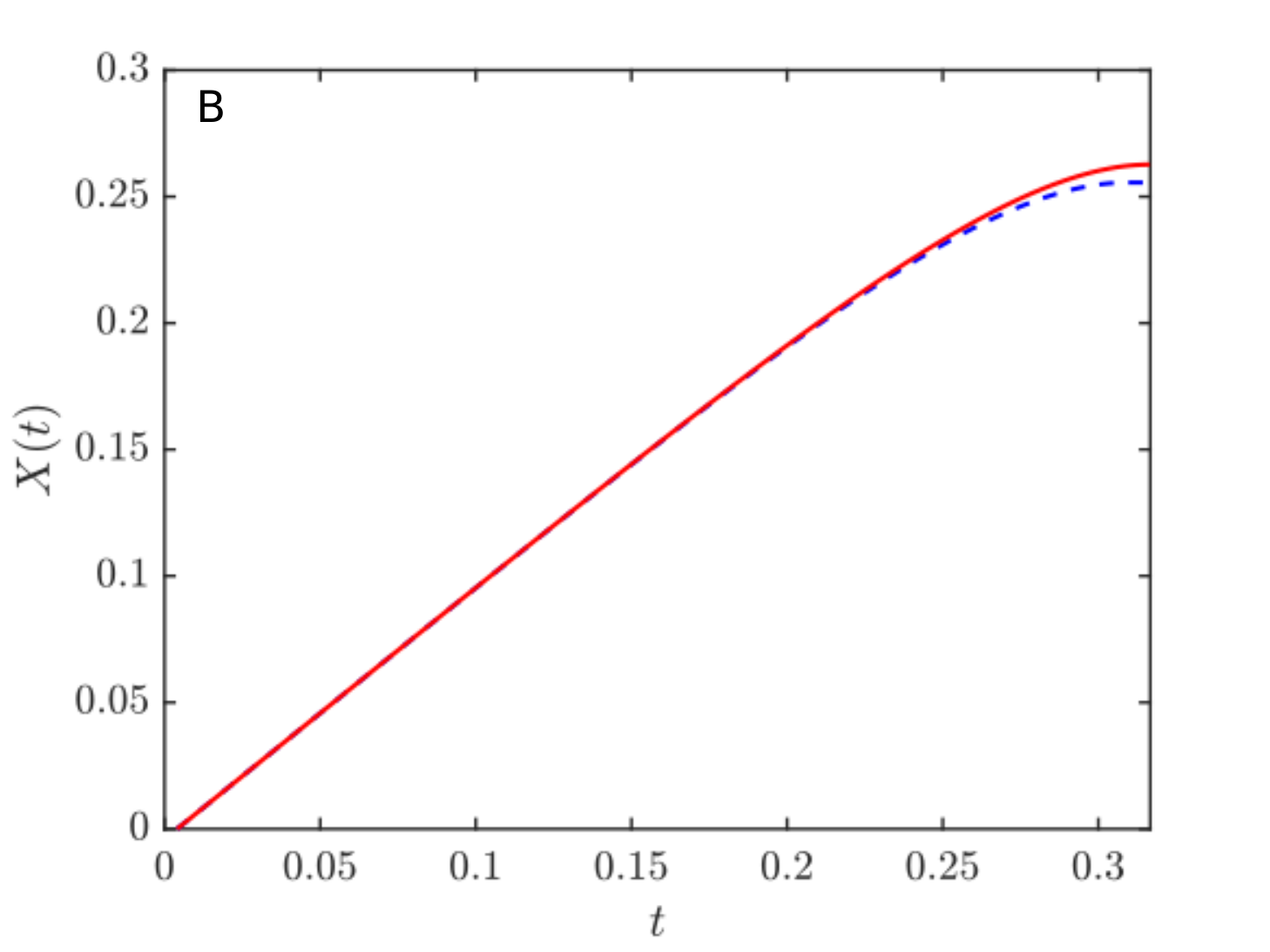}
\includegraphics[angle=0,width=0.5\textwidth]{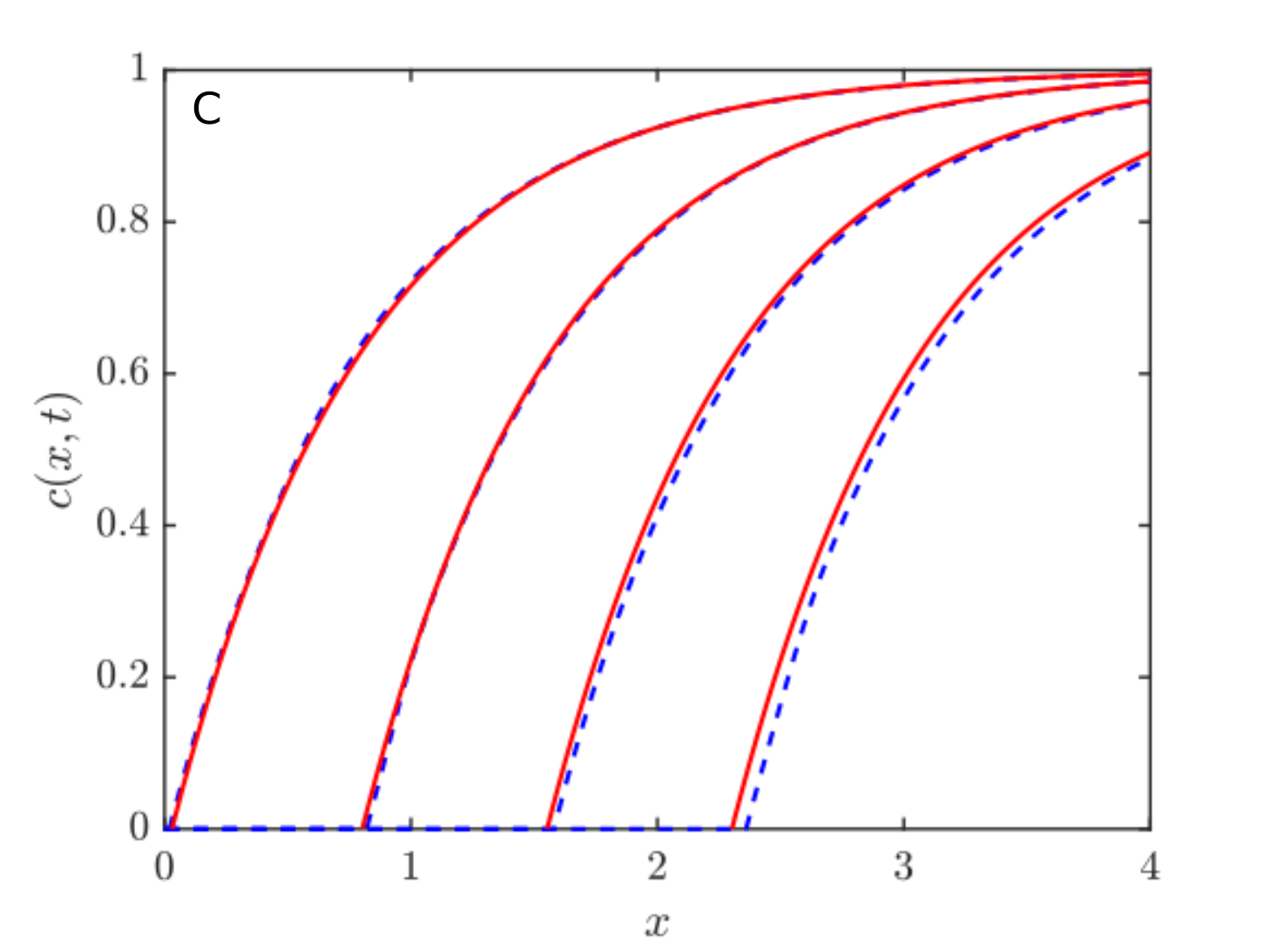}\includegraphics[angle=0,width=0.5\textwidth]{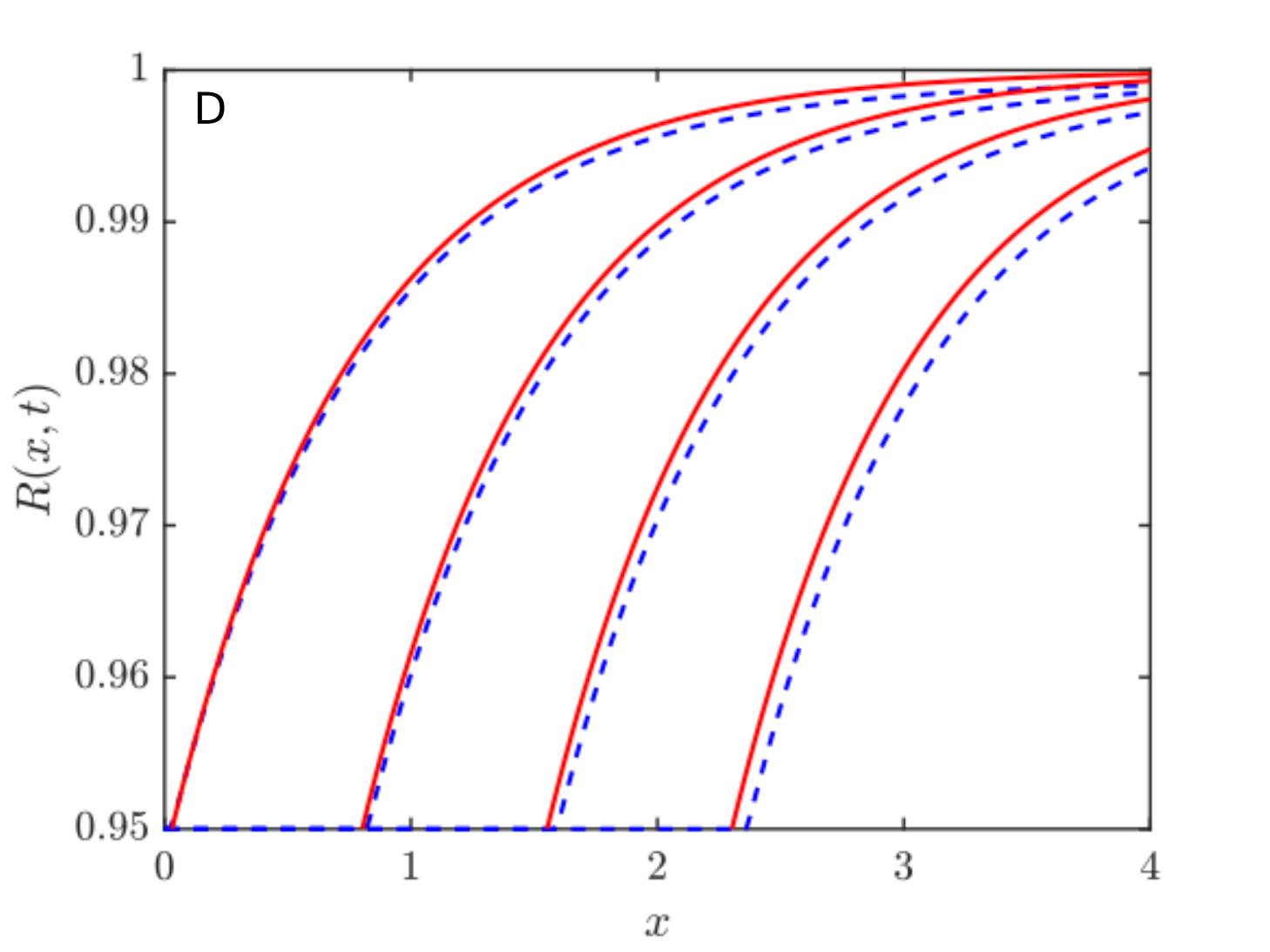}
\caption{Numerical (dashed) and perturbation (solid) solutions using $L=4$, $\Da=\Pe^{-1}=0.001$, $R_c=0.95$, $\epsilon_i=0.75$. Panel A shows the outlet concentration and radius, $c(L,t), R(L,t)$. The concentration profile in panel C and radius in Panel D correspond (from left to right) to times $t = 0.05, 0.1, 0.15, 0.2$. }
\label{fig_sec5_1}
\end{figure}

In Figure \ref{fig_sec5_1}  we compare the predictions of the numerical solution (dashed line) with that of the perturbation (solid line) at various positions and times, with $\Da=\Pe^{-1}= 0.001$, $\epsilon = 0.75$ and $R_c=0.95$. 
Figure \ref{fig_sec5_1}A shows the variation of $c, R$ at the outlet. For early times the fluid is saturated with material and $c(L,t) \approx 1$. This lasts until around $t \approx 0.06$ when $c$ starts to decrease monotonically until it reaches zero at $t \approx 0.32$. Since the saturated fluid cannot remove material $R \approx 1$ until the concentration begins to decrease after which it slowly decays to the core value $R_c=0.95$. The numerical and perturbation solutions are clearly close, with only slight differences visible at the end of the process.

Figure \ref{fig_sec5_1}B shows the variation of $X(t)$. The perturbation and numerical solutions are virtually indistinguishable except at the very end. Recalling the discussion of an approximately linear section in \S \ref{aproxS} for small $E_L$ here we note that $E_L = \exp (-L/\epsilon_i) = \exp(-4/0.75) \approx 0.005$. According to equations (\ref{Xe1}, \ref{Xl1}) the linear section of the curve is well approximated by $X(t) \approx t-t_1$, the exponential component only becoming noticeable in the final stages. 

The concentration and radius at times $t = 0.05, 0.1, 0.15, 0.2$ are shown in Figures \ref{fig_sec5_1}C,D. The self-similar form is evident and the numerical solution closely follows the travelling wave form of the perturbation. It may be noted that at the exit $x=L=4$ the concentration gradient appears to be non-zero, contradicting the applied boundary condition $c_x(L,t)=0$. This is due to the presence of a very narrow boundary layer of thickness $\ord{\sqrt{Pe^{-1}}}$. With the chosen value $Pe^{-1}  = 0.001$ the resultant boundary layer has thickness of order $\ord{0.03}$ which is hard to distinguish on the above graph. However, we have verified its existence in separate calculations using a much finer grid. 

\begin{figure}[htb]
\centering
\includegraphics[angle=0,width=0.5\textwidth]{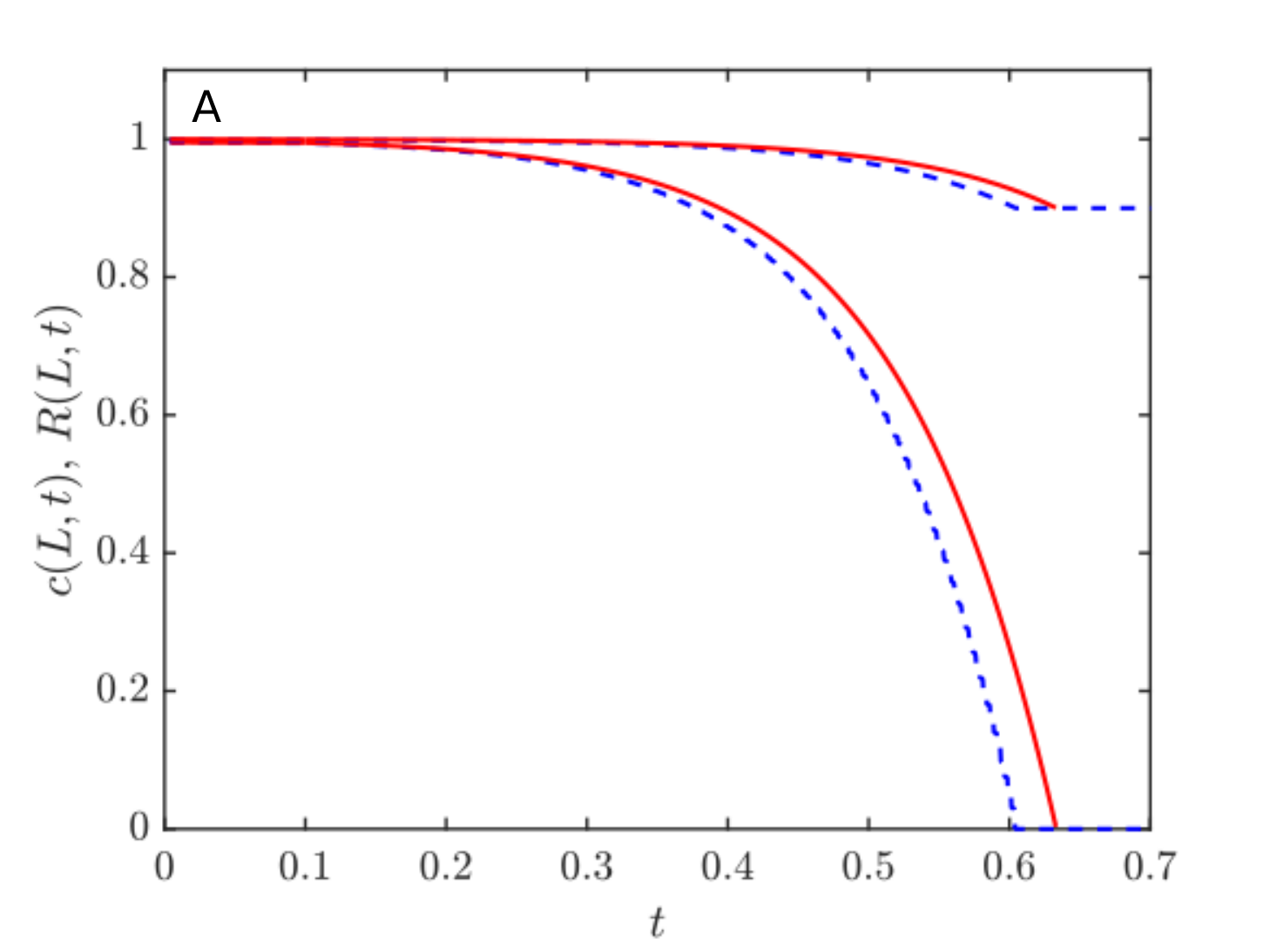}\includegraphics[angle=0,width=0.5\textwidth]{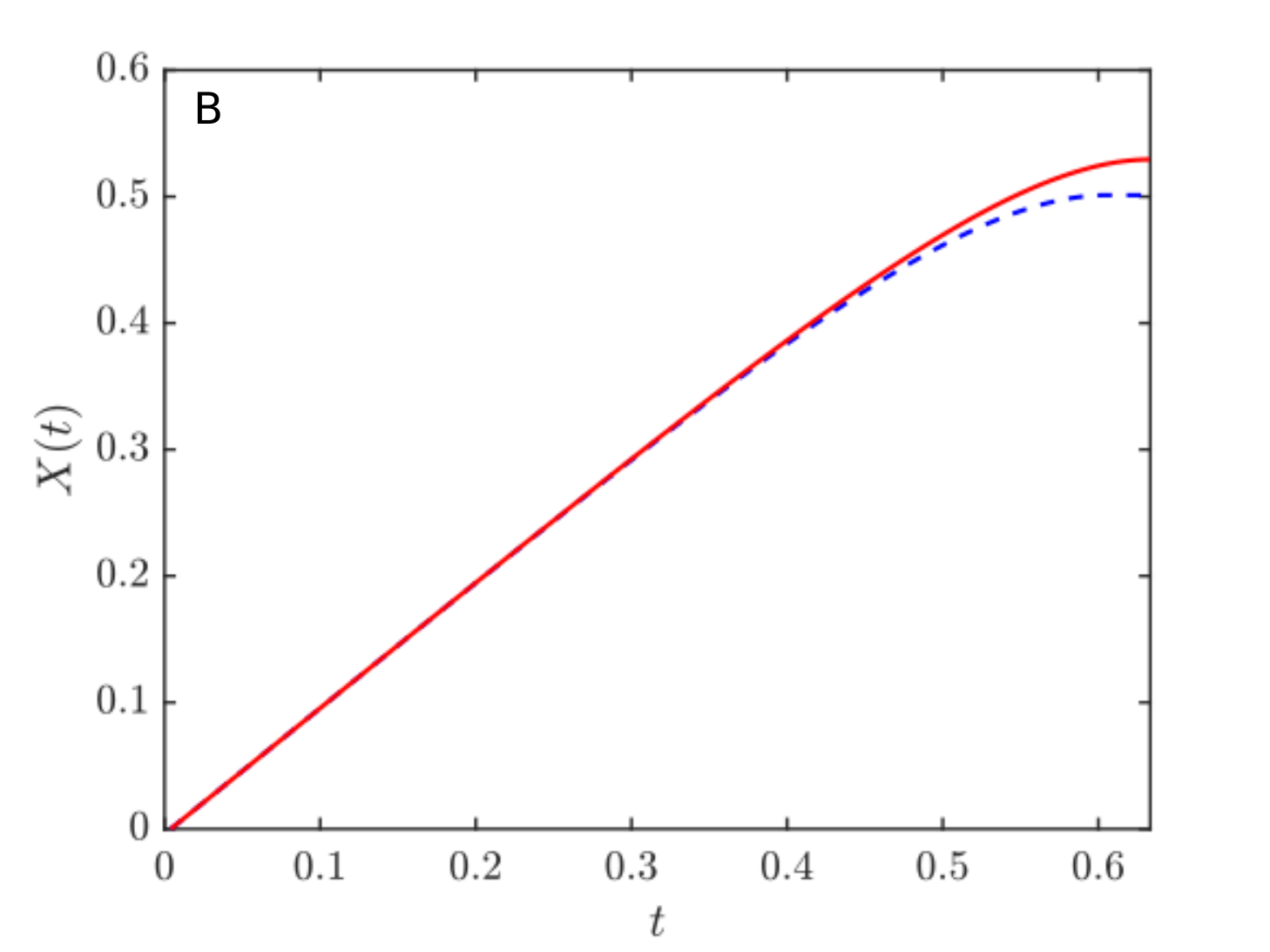}
\includegraphics[angle=0,width=0.5\textwidth]{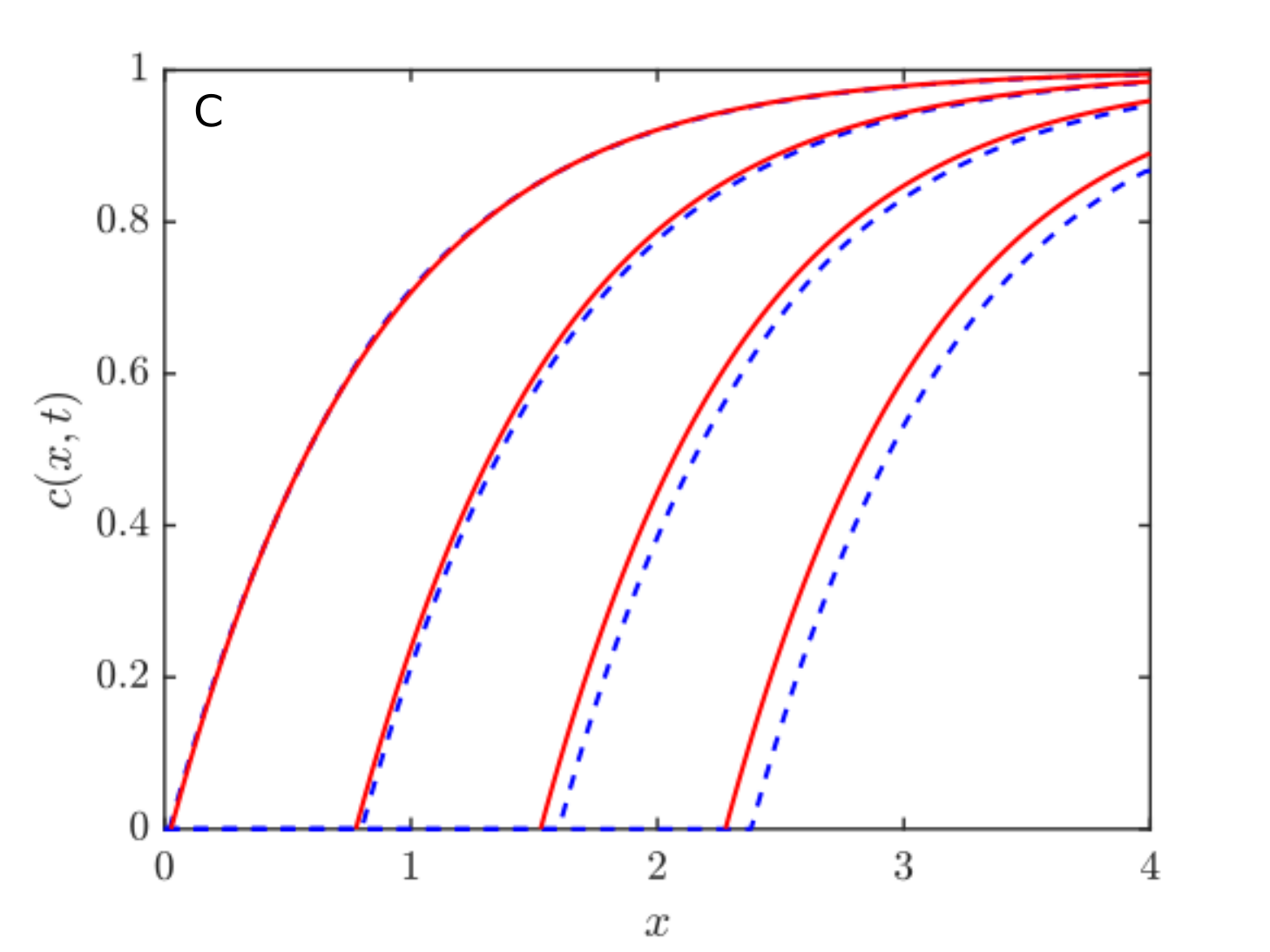}\includegraphics[angle=0,width=0.5\textwidth]{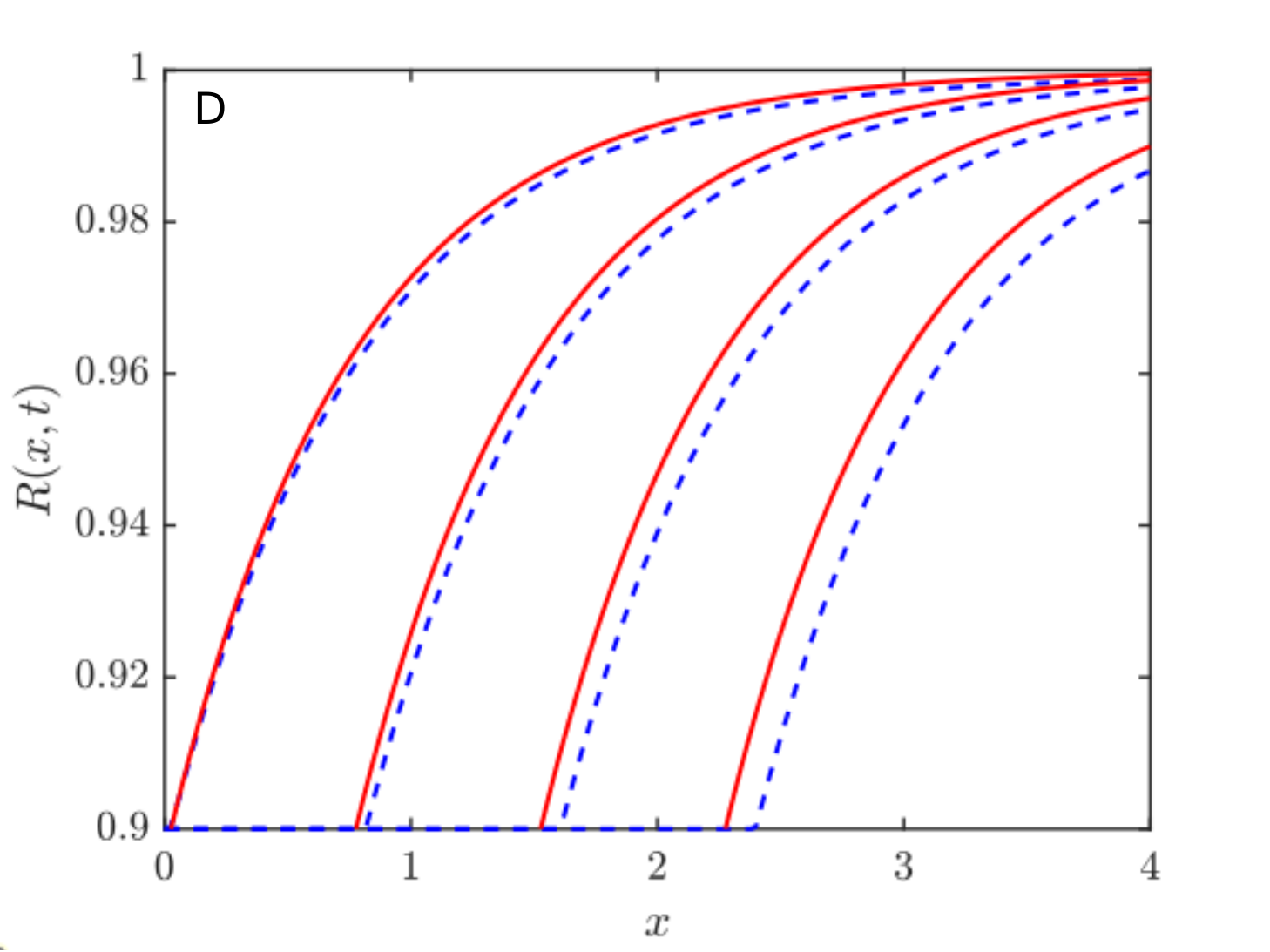}
\caption{Numerical (dashed) and perturbation (solid) solutions using $L=4$, $\Da=\Pe^{-1}=0.001$, $R_c=0.9$, $\epsilon_i=0.75$. The profiles in panels C, D  correspond (from left to right) to times $t = 0.1, 0.2, 0.3, 0.4$.}
\label{fig_sec5_2}
\end{figure}

Figure \ref{fig_sec5_2} shows a similar set of results but now $R_c=0.9$. The same general behaviour may be observed but with slightly larger differences between the two sets of curves. This is to be expected: the perturbation is based on the small parameter $\delta = 1-R_c$. As $R_c$ decreases $\delta$ increases and hence so does the error. However, again the two solutions for $X$ are indistinguishable until the final stages and have an approximately linear form for a large amount of the process. Separate numerical calculations have confirmed the existence of a boundary layer over which the concentration gradient tends to zero.

\begin{figure}[H]
\centering
\includegraphics[angle=0,width=0.5\textwidth]{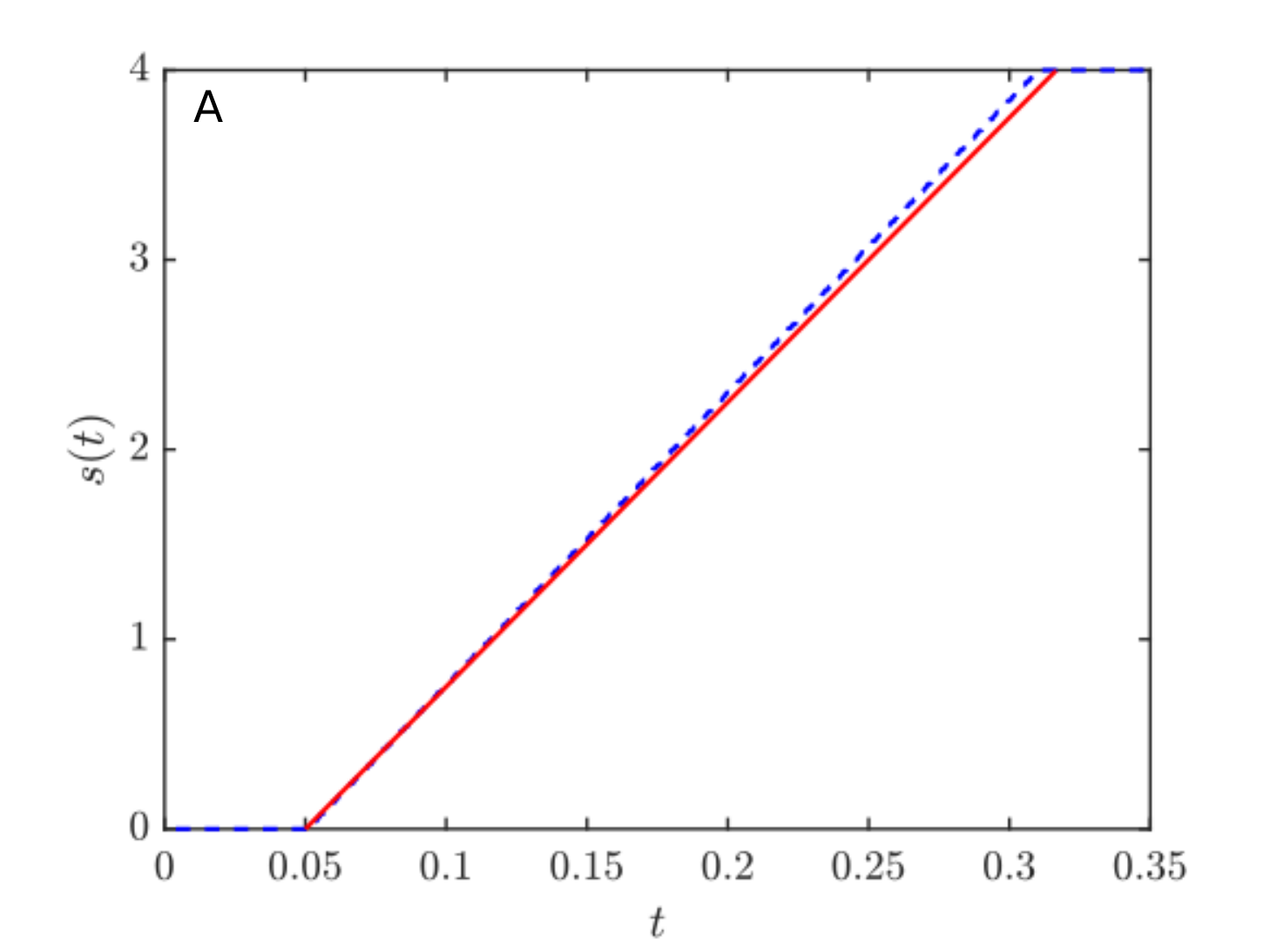}\includegraphics[angle=0,width=0.5\textwidth]{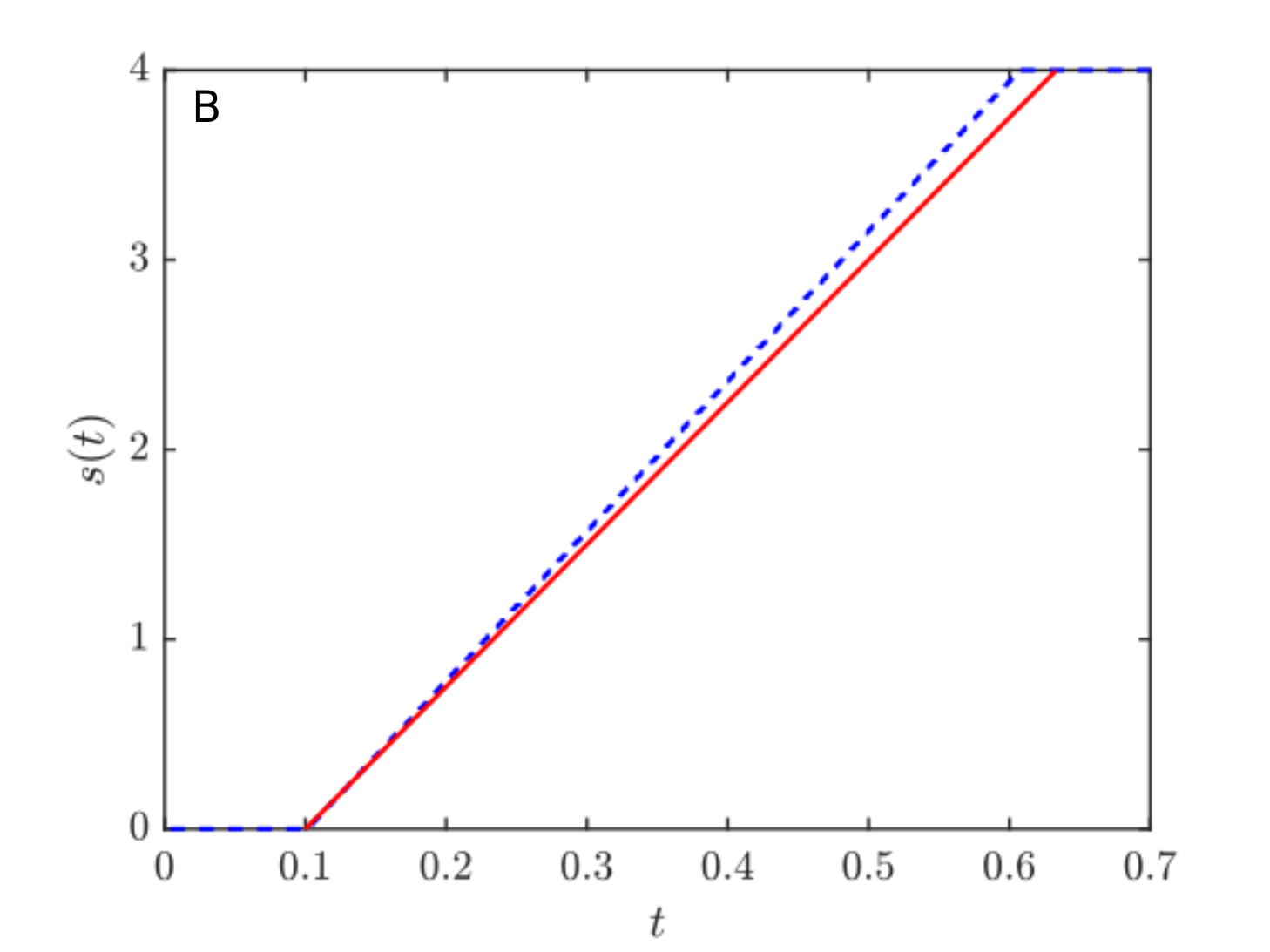}
\caption{Evolution of the moving boundary $s(t)$ from the numerical and perturbation solutions for $L=4$, $\Da=\Pe^{-1}=0.001$, $\epsilon_i=0.75$ and (A) $R_c=0.95$, (B) $R_c=0.9$. }
\label{fig_sec5_3}
\end{figure}

The main difference between numerics and perturbation in panels C and D in Figures \ref{fig_sec5_1}, \ref{fig_sec5_2} is really the point where the curve begins: the numerical solution is slightly ahead. In Figure \ref{fig_sec5_3} we compare the position of the front $s(t)$ for the two cases. The perturbation is always slightly behind, indicating a slightly slower velocity. This could be improved by taking higher order terms but  may be complicated by the contribution of $\Da/\delta, \Pe^{-1}$ terms which may enter into lower order expressions. As time increases the small difference in speeds then acts to move the curves farther apart. However, even at the end of the process, for the case shown in Figure \ref{fig_sec5_3}B the error is below 5\%. So we may conclude the perturbation solution is, in general, highly accurate. Further, since the extracted fraction is the quantity of main interest the very close correspondence between curves suggests that the perturbation is particularly accurate for tracking this quantity. Consequently in the following section, where we examine the removal of lanolin from woollen fibres, we will focus primarily on the perturbation solution since this gives a clear picture of the role of the system parameters.

\section{Application to lanolin removal from wool fibres}
\label{LanSec}
In this section we apply the above model to the removal of lanolin from wool fibres. For this we take the operating conditions and data from the experiments described in \cite{Eychenne01}. 

\begin{table}[H]
\caption{ Experimental operating conditions of \cite{Eychenne01}.}
\label{Table1}
\begin{center}
\begin{tabular}{ |c|c| }
\hline
 {Temperature} & 30 \units{o}C\\
\hline
 {Pressure} & 70, 120, 150 bar\\
\hline
 {Solvent mass flowrate, $\dot{m}^*$} & 3, 4, 5 kg/h\\
 \hline
 $\dot{m}^*$ (in SI units)  & 8.33, 11.11, 13.89 $\times 10^{-4}$ kg/s\\
\hline
 {Wool packing density, $\rho_{B}^*$} & 127, 159, 227, 318 kg/m\units{3}\\
\hline
 {Solvent composition, $\%$ wt.} & 95$\%$ CO$_\text{2}$ – 5$\%$ ethanol\\
\hline
\end{tabular}
\end{center}
\end{table}

\begin{table}[htb]
\caption{Parameter values from the experiments of \cite{Eychenne01}}
\label{Table2}
\begin{center}
\begin{tabular}{ |c|c|c|c|c|c|c|c| }
\hline
\textbf{Property} & \textbf{Sym.} & \textbf{Units} & \multicolumn{4}{|c|}{\textbf{Value}} & \textbf{Ref./Method}\\
\hline
\hline
\multicolumn{8}{|c|}{Extractor vessel and wool load}\\
\hline
Inner rad. & $R_{b}^*$ & m & \multicolumn{4}{|c|}{0.015} & \multirow{3}{*}{\cite{Eychenne01}}\\
\cline{1-7}
Section & $A_{b}^*$ & m\units{2} & \multicolumn{4}{|c|}{7.0686$\times$10\units{-4}} & \\
\cline{1-7}
Wool load & $m_{p}^*$ & kg & \multicolumn{4}{|c|}{0.013} & \\
\hline
Wool dens. & $\rho_{p}^*$ & kg/m\units{3} & \multicolumn{4}{|c|}{1314} & \cite{Simpson02} $\&$ \cite{King26}\\
\hline
Bulk dens. & $\rho_{B}^*$ & kg/m\units{3} & 127 & 159 & 227 & 318 & \cite{Eychenne01}\\
\hline
Volume & $V_{b}^*$ & m\units{3} & 1.0$\times$10\units{-4} & 8.2$\times$10\units{-5} & 5.7$\times$10\units{-5} & 4.1$\times$10\units{-5} & $V_{b}^*=m_{p}^*/\rho_{B}$\\
\hline
Length & $L^*$ & m & 0.145 & 0.116 & 0.081 & 0.058 & $L^*=V_{b}^*/A_{b}^*$\\
\hline
Porosity & $\epsilon_{i}$ & -- & 0.903 & 0.879 & 0.827 & 0.758 & $\epsilon_{i}=1-\rho_{B}^*/\rho_{p}^*$\\
\hline
\multicolumn{8}{|c|}{Wool fibres and lanolin}\\
\hline
Initial rad. & $R_{i}^*$ & m & \multicolumn{4}{|c|}{10\units{-5}} & \cite{Eychenne01}\\
\hline
$\%$ wt. lan. & \%lan &  & \multicolumn{4}{|c|}{9.6$\%$} & Estimated\\
\cline{1-7}
Layer rad. & $R_{w}^{*}$ & m & \multicolumn{4}{|c|}{9.52$\times$10\units{-6}} & from\\
\cline{1-7}
Core rad. & $R_{c}^{*}$ & m & \multicolumn{4}{|c|}{9.25$\times$10\units{-6}} & \cite{Eychenne01}\\
\hline
Density & $\rho_{e}^{*}$ & kg/m\units{3} & \multicolumn{4}{|c|}{940} & Aspen Plus v9\\
\hline
\multicolumn{8}{|c|}{Solvent}\\
\hline
Density & $\rho_{s}^{*}$ & kg/m\units{3} & \multicolumn{4}{|c|}{783.3} & \multirow{3}{*}{Aspen Plus v9}\\
\cline{1-7}
Critical T & $T_{crit}^*$ & K & \multicolumn{4}{|c|}{310.55} &\\
\cline{1-7}
Critical P & $P_{crit}^*$ & Pa & \multicolumn{4}{|c|}{7.73$\times$10\units{6}} &\\
\hline
\multirow{2}{*}{Viscosity} & \multirow{2}{*}{$\mu^{*}$} & \multirow{2}{*}{Pas} & \multicolumn{4}{|c|}{\multirow{2}{*}{6.48$\times$10\units{-5}}} & Aspen Plus v9\\
& & & \multicolumn{4}{|c|}{} & with \cite{Fields11}\units{1}\\
\hline
\end{tabular}
\end{center}
\footnotesize{\units{1}Fields et al. \cite{Fields11} show that at 37 \units{o}C, the viscosity of the mixture 95$\%$ CO$_\text{2}$:5$\%$ ethanol doesn't change significantly with pressure. In their work, the viscosity of the solvent at 10\units{7} Pa is 6.35$\times$10\units{-5} Pas.}
\end{table}

In Table \ref{Table1} we present the operating conditions, in Table \ref{Table2} we present the appropriate parameter values. The  extractor vessel was a stainless steel AISI 316L Separex SCF 200. According to \cite{Eychenne01} the initial wool load consisted of 60-65\%  wool fibres, 10-15$\%$ wax (lanolin) and proteins, 10\% soluble stains (salts), 1-20\% soil and vegetable matter (since the total amount with 1\% of soil and vegetable matter does not reach 100\%, we assume this is an error and should read 10\%). The key parameters not provided in the Tables are the mass transfer coefficient $k^*$ and the saturation concentrations $c_s^*, c_w^*$ which are {\it a priori} unknown. They vary with the operating conditions and may be  determined through comparison with experimental data. However, to determine the order of magnitude of various non-dimensional terms we note that in \cite{Valv19} for the experiments at 120 bar, the values given are $k^* = 4.1 \times 10^{-6}$s$^{-1}$ and $c^*_s  \approx 0.632$kg/m$^3$ (the value actually quoted is
0.807 g/kg solvent, multiplying by the solvent density $\rho_s$ and converting to kilograms we obtain $c_s^* = 0.807 \times 10^{-3} \times 783.3 = 0.632$kg/m$^3$). 
To estimate the diffusion coefficient we refer to \cite[Fig. 5]{Abaroudi99}, this shows that for a supercritical liquid the inverse Bodenstein number $D_i^*/(2 u^* R^*) \approx 2$ is approximately constant for Reynold's numbers $\mathrm{Re}\in [0.1, 20]$ (note we have converted their definition from superficial velocity and particle diameter). This indicates that $D_i^* \approx 4 u^* R^* = 4 \epsilon_i u^*_i R^*/\epsilon$ varies with $\epsilon$ and $R^*$.  Using the average of the mass flow rates from Table \ref{Table1},  $\dot{m}^* \approx  1.1 \times 10^{-3}$kg/s, and the average void fraction $\epsilon_i = 0.83$ of Table \ref{Table2} we obtain a representative initial interstitial velocity
\bea
u^*_i = \frac{\dot{m}^*}{\epsilon_i \rho_s^* \pi R_b^{*2}} = 0.0024 \mathrm{m/s} \, .
\eea
With a fibre radius $R^* = 10 \mu$m we find $D_i^*= 9.6 \times 10^{-8}$m$^2$/s. We may now also calculate typical length- and time-scales of the extraction process
\bea
{\cal L}^* =    \frac{R_{i}^* u_i^*}{2 (1-\epsilon_i) k^*} \approx 0.017 \mathrm{m} \, ,\quad  \tau^* = \frac{\rho_e^* R_{i}^*}{k^*c_s^*} \approx 3753 \mathrm{s}\, .
\eea
These indicate that the width of the main extraction zone is of the order 2cm while the time-scale for the process is of the order of one hour. 
The non-dimensional parameters may now be evaluated
\bea
\label{DaPeData}
\Da= \frac{{\cal L}^*}{u^*_i \tau^*} \approx 0.002 \, , \quad \Pe^{-1} = \frac{4 R_i^*}{{\cal L}^*} \approx 0.002 \, , \quad \delta = 0.075 \, .
\eea

In terms of the perturbation solution we note that the error incurred by neglecting $\Da, \Pe^{-1}$ will then be of order 0.2\%, so justifying  their omission from the respective equations. However, in the analysis we actually rescale time with $\delta$, meaning that the time derivative term is  of order $\Da/\delta = 0.002/0.075= 0.03$. So it is still valid to neglect the time derivative, even with this re-scaled  time variable, but now the associated errors are of order 3\%.

The exponential coefficient $E_L = \exp(-L^*/(\epsilon_i {\cal L}^*))  \sim \exp(-0.1/(0.83 \times 0.017) \approx 10^{-3}$ is smaller than neglected terms and so  has very little influence on the results.

\subsection{Results and discussion}

To determine the unknown parameter values we take the experimental data  of \cite{Eychenne01} and carry out an optimisation procedure. The experiments indicate a clear change in solubility, see  the discussion in \cite{Valv19}, consequently we seek three unknowns $k^*, c_s^*, c_w^*$. Since $c_s^*$ and $c_w^*$ are thermodynamic properties, it is sensible to consider them as only dependent on temperature and pressure. Since the experiments are isothermal we anticipate a unique $c_s^*$ and $c_w^*$ for each pressure. The mass transfer coefficient, $k^*$, clearly depends on the flow (the force of  which depends, for example, on the fluid velocity and also its path through the material) and so will vary throughout the experiments. Consequently this is calculated for each data set. We employ nine data sets from \cite{Eychenne01} which involve three different pressures, leading to a total of fifteen parameters to optimize, i.e. nine mass transfer coefficients, three first fraction and three second fraction solubilities. In order to start our procedure initial guesses are required. Approximate values of the first two were discussed earlier, the final unknown $c_w^*$ we take as some fraction of $c_s^*$ (in practice we set $c_w^*=c_s^*/10)$. 
The optimisation procedure employs an interior-point algorithm, using the MATLAB package \textit{GlobalSearch} combined with the local solver \textit{fmincon},  in order to reach a global constrained optimum. The objective function takes the form of the sum of the absolute value of the errors,
\bea
f=\sum_{i=1}^n\mid X^{*exp}_{i}-X^{*}_{i}\mid \, ,
\eea
where $X_i^*$ comes from the perturbation solution and $X^{*exp}_{i}$ are the values provided by \cite{Eychenne01}.

For the two solubility problem $X_i^*$ has four components defined by equations (\ref{Xestwosol} - \ref{Xf2}). The switch between each stage depends on the value of the unknowns, consequently initially the whole $X$ curve is calculated based on the estimated values and the objective function is then evaluated. The optimization algorithm is applied and new values for the parameters are calculated. These are used to recalculate the switching times and the process repeated until convergence is achieved. 
In order to compare model predictions with the single solubility model 
we carry out the  procedure a second time but using only two unknowns  $k^*, c_s^*$.

\begin{figure}[htb]
  \centering
  \includegraphics[scale=0.4]{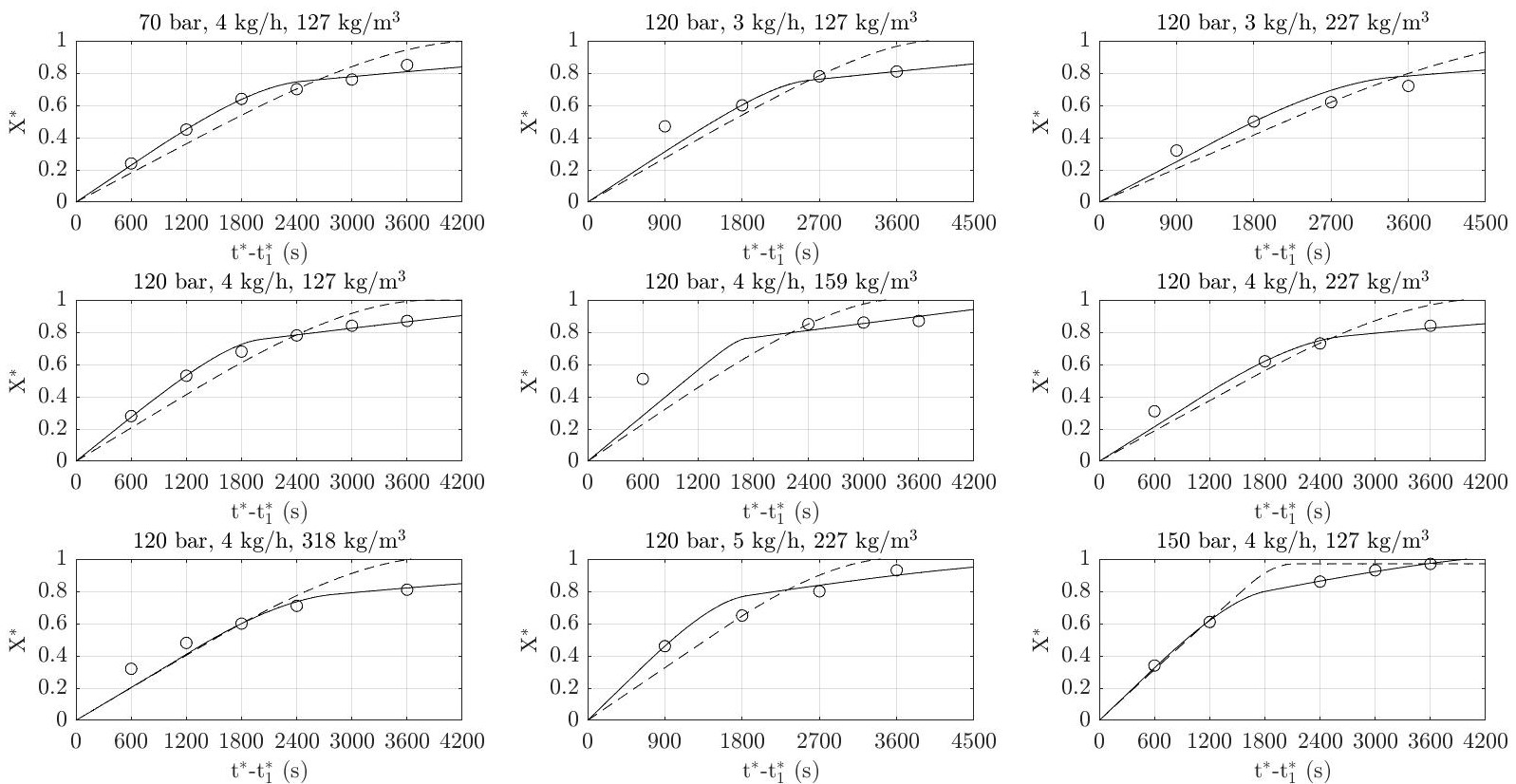}
  \caption{Extracted fraction curves obtained after optimizing the parameters $k^*$, $c_s^*$ and $c_w^*$, using the perturbation method. The solid line represents the  two solubility model, the dashed line the single solubility model and circles correspond to experimental data points. }
  \label{AllX}
\end{figure}

In Figure \ref{AllX} we present a comparison of results from the single and double solubility perturbation solutions and the experimental data of \cite{Eychenne01} for the extracted fraction at various pressure, flow rate and wool packing conditions. Circles represent the data points, solid lines the two solubility model and dashed lines the single solubility model. 
In general the agreement between experiment and the two solubility model is excellent.  However there are two rather prominent points where the agreement is clearly not so good, these are the first data points of the $P=120$ bar, 3 kg/h, 127 kg/m\units{3} and  4 kg/h, 159 kg/m\units{3} graphs. In both cases the graphs only have four data points and only a single data point within the first linear stage. Consequently we may assume that in these cases there is not sufficient early time data to accurately characterise the behaviour. The single solubility model is clearly nowhere near as accurate, it suffers from the attempt to fit all data points. The inability to match the data confirms the observation of previous studies that lanolin removal occurs in two stages.
The single solubility model could easily be adjusted to match just the early data points, where it is valid. 

In \S \ref{aproxS} we discussed the linear forms of the extraction curve. This is apparent in the figures: the two solubility model clearly shows two linear sections while the one solubility case has a single linear section. 
Given the approximately linear behaviour for early times in both cases it is a simple matter  to calculate  $c_s^*$ from the formula $X^* \approx \dot{m}^* c_s^* (t^*-t_1^*)/(M_{tot}^* \rho_s^*)$ (i.e. the gradient is $\dot{m}^* c_s^* /(M_{tot}^* \rho_s^*)$). This could be achieved  using only the first data point for $X^*$ or more early data points (if available) and then averaging.
Similarly we could estimate $c_w^*$ through the gradient of the second linear stage $\dot{m}^* c^*_w/(M_{tot}^* \rho_s^*)$.

The early time solutions for $X_e^*$, (equations \eqref{Xesingle} and \eqref{Xestwosol}), show a linear variation with time and exhibit only a very weak dependence on $k^*$ (through the value of $E_L$, which is of order $10^{-3}$). Consequently, the early time solution, which in Fig. \ref{AllX} lasts for almost half of the process time,
indicates that $X^*$ is independent of $k^*$ with a high degree of accuracy. This phase corresponds to the period when the exit concentration $c^* \approx c_s^*$ (with errors of the order $E_L$, i.e. approximately 0.1\%).  The second linear stage also reveals little information about $k^*$. This also corresponds to a saturated outlet fluid but this time corresponding to the saturation value of the second layer, i.e.  $c^* \approx c_w^*$. The value of $k^*$ can therefore only be determined reliably through knowledge of the transition zone between the two linear sections or the very late time data (beyond the second linear section).
The physical interpretation is that if the fluid is saturated at the outlet then the outlet measurements of the extracted fraction alone cannot be used to determine the mass transfer rate. The fluid could have been saturated close to the inlet or just before the outlet, there is no reliable way to distinguish between the two if the only data is values of $X^*$.

The parameter values  for $k^*$, $c_s^*$ and $c_w^*$, obtained through the optimisation of the two solubility model are presented in Table \ref{tab:Xopt}. 
The value of $c_s^*$ obtained from the  formula 
\bea
c_s^* \approx M_{tot}^* \rho_s^* X^*/(\dot{m}^*  (t^*-t_1^*)) ~ ,
\eea
using only the first data point is also given and  labelled ``$c_s^*$ 1\units{st} point". Overall the agreement between the two $c_s^*$ values is good, suggesting that using just the first data point is sufficient to provide a rough estimate of the value for the  solubility (provided that point occurs within the first stage). Examination of the "$c_s^*$ 1\units{st} point" row corresponding to $P=120$ bar indicates that the majority of values are close to 0.45 kg/m\units{3}. The exceptions are the two cases mentioned earlier, for 3 kg/h, 127 kg/m\units{3} and  4 kg/h, 159 kg/m\units{3}, which have only a single data point within the linear region. Ignoring these two cases the remaining values show a maximum of around +11\%/-13\% deviation from 0.45. The range of $k^*$ values in the table, $[0.83,2.47] \times 10^{-6}$m/s, are consistent with those of Figure 12 of Puiggen\'e et al. \cite{Puiggene97} and also the low Reynolds number correlation reported by Tan et al. \cite{Tan88} for solid-SCF systems. They are close to those reported by Valverde et al. \cite{Valv19} (although they use a slightly less accurate value for $R_c^*$).

\begin{table}[H]
\caption[tiny]{\label{tab:Xopt} Value of the parameters obtained from optimization procedure for extraction at 30 \units{o}C and diverse pressure, mass flow rate and packing density conditions. }
\begin{center}
\begin{tabular}{ |c|c|c|c|c|c|c|c|c|c| }
\hline
P (bar) & 70 & \multicolumn{7}{|c|}{120} & 150\\
\hline
$\dot{m}^{*}$ (kg/hr) & 4 & \multicolumn{2}{|c|}{3} & \multicolumn{4}{|c|}{4} & 5 & 4\\
\hline
$\rho_{B}^{*}$ (kg/m\units{3}) & 127 & 127 & 227 & 127 & 159 & 227 & 318 & 227 & 127\\
\hline
$k^{*}$ ($\times$10\units{-6}) (m/s) & 1.63 & 2.30 & 0.83 & 2.47 & 0.92 & 1.01 & 0.92 & 1.88 & 1.67\\
\hline
$c_s^*$ 1\units{st} pt (kg/m\units{3}) & 0.38 & 0.66 & 0.45 & 0.44 & 0.80 & 0.49 & 0.50 & 0.39 & 0.54\\
\hline
$c_s^*$ (kg/m\units{3}) & 0.40 & \multicolumn{7}{|c|}{0.45} & 0.57\\
\hline
$c_w^*$ (kg/m\units{3}) & 0.056 & \multicolumn{7}{|c|}{0.069} & 0.116\\
\hline
\end{tabular}
\end{center}
\end{table}

Physically we expect the value of $k^*$ to increase with an increase in flow rate, $\dot{m}^*$ (due to the  increase in the force of the liquid on the solid). It could also be expected to vary with the bulk density but in a less obvious way. (The bulk density affects the void fraction $\epsilon_i$ which in turn affects the interstitial velocity, a low $\rho_B^*$ indicates low $u^* \sim 1/\epsilon_i$. However $k^*$  appears in the switching times and also $E_L = \exp(-L^*/(\epsilon_i {\cal L}^*))$. This combination  suggests a nonlinear response).  A nonlinear dependence is consistent with the results of \cite[Fig. 7]{Eychenne01} who report a maximum mass transfer rate at around $\rho_B^* = 100$ kg/m$^3$.

\begin{figure}[H]
  \centering
  \includegraphics[scale=0.4]{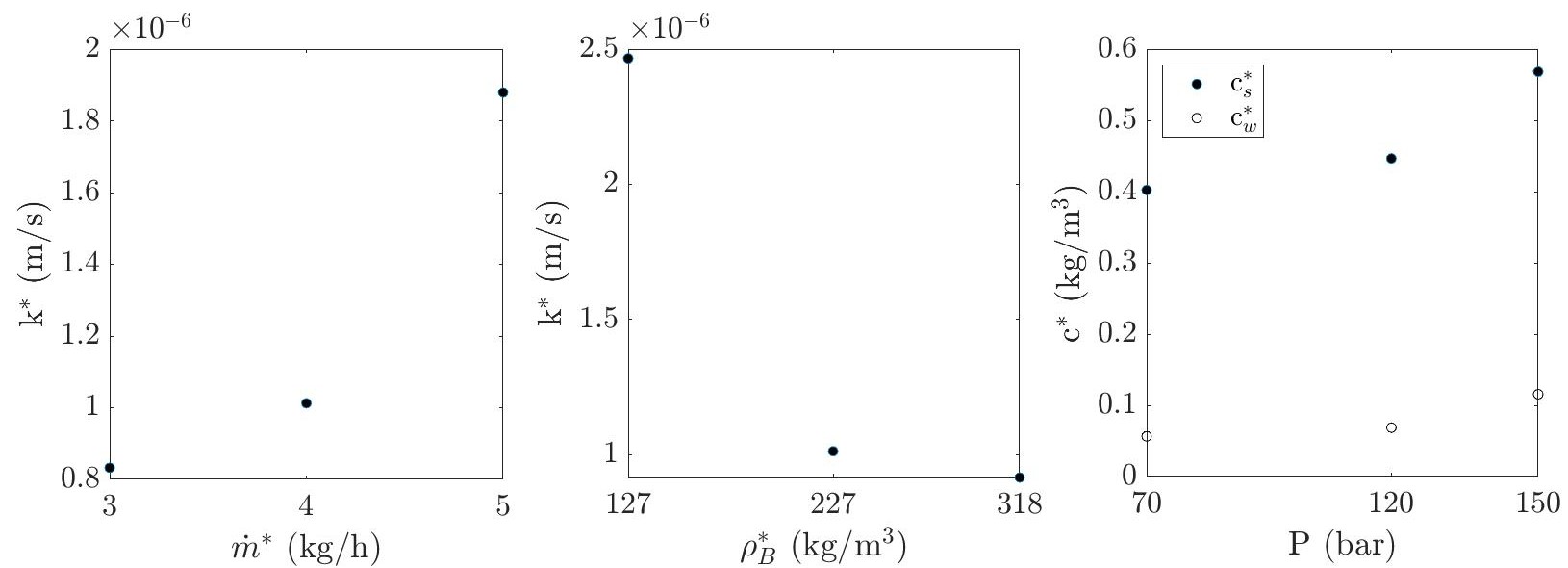}
  \caption{Dependence of k$^{*}$, $c_s^*$ and $c_w^*$ on diverse operating conditions at 30 \units{o}C. From left to right: $k^{*}$ versus $\dot{m}^*$ at 120 bar and 227 kg/m$^3$; $k^{*}$ versus $\rho_B^*$ at 120 bar and 4 kg/h; $c_{s}^{*}$ (solid points) and $c_{w}^{*}$ (void points) versus pressure.}
  \label{Compare}
\end{figure}

The influence of operating conditions on  $k^{*}$, $c_s^*$ and $c_w^*$ is shown in  Fig. \ref{Compare}. The first figure shows the variation of $k^*$ with $\dot{m}^*$ for fixed pressure and flow rate. As anticipated the value increases with increasing mass flow rate. When plotted against packing density, a monotonic decrease is observed. This behaviour is consistent with the \cite[Fig. 7]{Eychenne01} who report a maximum $k^*$ at $\rho_B^*>100$ kg/m\units{3} $k^*$ followed by a  monotonic decrease.
Our results show no clear dependency of $k^*$ on pressure, so we omit this figure (but note Valverde et al. \cite{Valv19} report a decrease of the mass transfer parameter with increasing pressure). Eychenne et al. \cite{Eychenne01} reported a yield increase with increasing pressure, but since solubility increases with pressure, it is difficult to determine the evolution of $k^*$ based on this set of experimental results.

Finally, the remaining plot in Figure \ref{Compare} shows the evolution of the two solubilities with pressure. Both fraction solubilities show a clear increase with increasing pressure, although it is more pronounced for $c_s^*$. This result is consistent with those of \cite{Eychenne01}, which suggest a significant increase of the solubility between 120 and 150 bar.
Note that the anomalous values obtained in Table \ref{tab:Xopt} at 120 bar, 4 kg/h and 159 kg/m\units{3} are excluded from the graphs. . 

\begin{figure}[htb]
  \centering
  \includegraphics[scale=0.35]{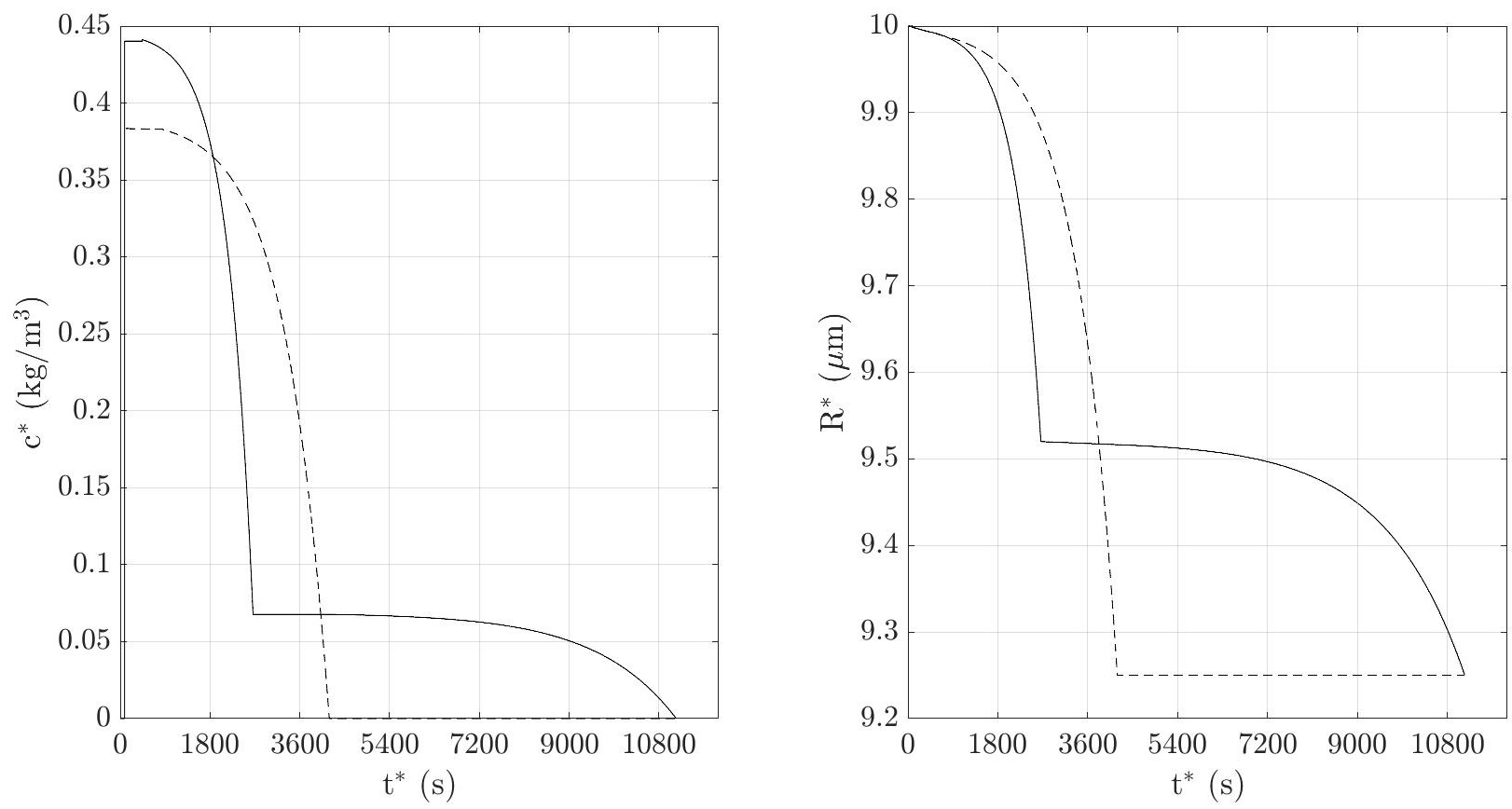}
  \caption{Concentration and radius profiles obtained from the perturbation method at  120 bar, 3 kg/h and 127 kg/m\units{3}. The solid line represents  the two solubility model,  the dashed line is the single solubility model.}
  \label{CR2sol1203127}
\end{figure}

Figure \ref{CR2sol1203127} demonstrates how the outlet concentration and radius evolve according to the perturbation solutions, using the parameter values obtained through the optimisation.  The single solubility model is shown as the dashed line. The initial value of 
the outlet concentration is zero until time $t_1^*$, when it suddenly jumps up. Since $t_1^*$ is a lot less than the process time it is difficult to see. This is followed by a period of very slow decrease between $t_1^*, t_2^*$, from \eqref{cRStage2} it may be observed that the gradient is proportional to $E_L$. There is a slight mismatch between the solutions at $t_2^*$ (a consequence of neglecting the time derivative which, as discussed, can lead to errors of the order 3\%) this is followed by a rapid, nonlinear decrease to zero, where all material has been extracted after around 4000s. The two solubility model has similar qualitative features, a rapid jump followed by (a much shorter) linear decrease, a slight mismatch in solutions and then two stages of nonlinear decrease. The second nonlinear stage shows a very slow decrease due to the low value of $c_w^* < c_s^*$. The process ends at around 10800s. In the second figure, for the one solubility model, after time $t_1^*$ the radius decreases monotonically to the final value of $R_c^* = 9.25 \mu$m. With the two solubility model the radius at first decreases more rapidly followed by a slow decay to the final value.

\section{Conclusions}

In this paper we have developed a novel, mathematically consistent one-dimensional model for the slow extraction of material from a porous matrix by a flowing fluid. We have also developed the first approximate analytical solution for this process. The analysis leads to expressions for the concentration and radius throughout the column. Integrating the expression for the outlet concentration provides an expression for the variation of extracted fraction with time. This is an important quantity since one of the main goals in the field is to efficiently extract material, also this is what is typically measured during experiments.  

The beauty of analytical solutions is that they clearly demonstrate the dependence upon the operating conditions, thus indicating how to optimise the extraction process. The dependence is not apparent with a purely numerical study. Here we found that for a significant portion of the process the extracted fraction  exhibits very simple behaviour, varying linearly with mass flow rate and the saturation solubility and decreasing linearly with the solvent density. In the final stages the response is nonlinear, involving a more complex dependence on operational parameters. 

The key model unknowns are the saturation solubility (or solubilities) and the mass transfer coefficient. In the case of a single solubility material the solubility is easily determined through matching the early data points for the extracted fraction with the analytical solution. In our model \lq early\rq \, corresponds to an extended  period where the outlet fluid is approximately saturated. The solution clearly shows that during the linear stage {\it it is not possible to accurately calculate the mass transfer coefficient}. The mass transfer coefficient may only be estimated through later data, when the outlet concentration starts to decrease. Consequently, in order to characterise an experiment it is important to have both early time data points to accurately determine the solubility as well as late data points, where the fluid is no longer saturated, to determine the 
mass transfer coefficient.

To validate the analytical solution comparison was made against a numerical solution. Subsequently the model was compared with experimental data for the extracted fraction of lanolin by a supercritical fluid. It has previously been established that the solubility of lanolin switches during the process, which required an extension of the solution to deal with a two solubility material. The analytical solution in this case involved three unknowns, the two solubilities and the mass transfer  coefficient. The analysis was tailored for the specific case of lanolin removal, where the first stages involve a single solubility followed by a period when the inlet region has been stripped to reveal the next layer, which is significantly more difficult to remove. 
During the initial stage, where the outer layer is being eroded, the model is identical to the single solubility  case. The two solubility model exhibited two clear linear regions, the first with gradient proportional to the first solubility while the second is proportional to the second solubility. This provides a way to quickly estimate both quantities. However, the solutions show that the mass transfer coefficient may only be reliably calculated using data from the transition between the two
linear regions or very late time data beyond the second linear period. 
This information is essential when interpreting experimental data.

There exist other possible scenarios involving, for example, a second layer with a lower solubility, a change in mass transfer rate or a constant radius. All of these could occur in different extraction processes. Since they were not relevant to the present study we did not consider all possible cases but they could be analysed using the methodology presented here and could form the basis of a future study.

\section*{Author Statement}
The work has not been published previously and it is not under consideration for publication elsewhere. Its publication is approved by all authors. If accepted, it will not be published elsewhere in the same form, in English or in any other language, including electronically without the written consent of the copyright-holder.

\section*{Acknowledgements}
T. Myers, F. Font and M. Aguareles acknowledge the support grant No. PID2020-115023RB-I00 financed by MCIN/AEI/10.13039/501100011033/ and by “ERDF A way of making Europe”. M. Aguareles  also acknowledges grant no.  MTM2017-84214-C2-2-P. F. Font acknowledges financial support from the Juan de la Cierva programme (grant IJC2018-038463-I). F. Font and T.G. Myers thank the CERCA Programme of the Generalitat de Catalunya, their work was also supported by the Spanish State Research Agency, through the Severo Ochoa and Maria de Maeztu Program for Centers and Units of Excellence in R\&D (CEX2020-001084-M). F. Font is a Serra H\'{u}nter Fellow.

\appendix

\section{Approximate solutions}
\label{aprox}
Here we provide the details of the approximate solutions described in Section \S\ref{aproxS}, for the one and two  solubility models.

\subsection{Single solubility model}\label{AppA1}\label{aprox1}
We start by defining $\delta=1-R_c\ll 1$ and rewrite $R=1-\delta \bar{R}$. In non-dimensional form $R \in [R_c, 1]$ and so $\bar{R} \in [0,1]$. Equation \eqref{ReqnND} becomes
\bea
\delta\pad{\bar{R}}{t} = 1-c\label{ReqnPerA}.
\eea
Equation \eqref{ReqnPerA}  indicates that to leading order $c=1$. which would correspond to a situation where the fluid has already extracted all the material. This is clearly not the situation when the process starts but we recall that the time scale was chosen for a thickness $R_i$ to be removed. In this study only a fraction $\delta$ is removed, which obviously takes a fraction $\delta$ of the time.  Consequently we must re-scale time, $\delta \bar{t} = t$, to correctly reflect this, obtaining
\bea
&&\pad{\bar{R}}{\bar{t}}  = 1-c\label{ReqnPerB}\,,\\
&&\frac{\Da}{\delta} \pad{}{\bar{t}}(\epsilon c) + \epsilon_i\pad{c}{x} = \Pe^{-1} \pad{ }{x}\left( D \pad{}{x}\big(\epsilon c\big)\right) +  (1-\delta\bar{R})(1-c)\label{ceqnPerB}\, ,\\ &&\epsilon =1-(1-\epsilon_i)(1-\delta\bar{R})^2=\epsilon_i+2\delta(1-\epsilon_i)\bar{R}-(1-\epsilon_i)\delta^2\bar{R}^2\,.\label{EpsEqnPer}
\eea

\paragraph{Stage 1}
In order to obtain an approximate solution we look for $\bar{R}=R_1+\ord{\delta}$ (i.e. $R = 1-\delta R_1 + \ord{\delta^2}$), $c=c_0+\delta c_1 +\ord{\delta^2}$, so then $\epsilon\sim \epsilon_i+2\delta(1-\epsilon_i)R_1+\ord{\delta^2}$. Taking into account that $\Da,\Pe^{-1}\ll \delta$ and balancing the order one and order $\delta$ terms in the mass balance and radius equations we obtain
\bea
\label{c0R1c1}
 \epsilon_i \pad{c_0}{x} =  1-c_0 \, , \qquad   \pad{R_1}{\bar{t}} =   1-c_0 \, , \qquad \epsilon_i \pad{c_1}{x} = -R_1 (1-c_0) - c_1
\, .
\eea
These are subject to
\bea
c_0(0,\bar{t}) = c_1(0,\bar{t})=R_1(x,0)=0 \, ,
\eea
such that
\bea
\label{ceeqn}
c_0(x,\bar{t}) = 1-e^{-x/\epsilon_i} \, , \qquad R_1(x,\bar{t}) =\bar{t} e^{-x/\epsilon_i} \, , \qquad c_1(x,\bar{t}) = (e^{-x/\epsilon_i}-1) \bar{t} e^{-x/\epsilon_i} \, .
\eea
Therefore, the approximate solutions in terms of $t=\delta\bar{t}$ read
\bea
R(x,t) &\sim&   1- \delta R_1(x,t) =   1-t e^{-x/\epsilon_i} \, ,\label{RApr1}\\
c (x,t) &\sim& c_0(x,t) + \delta c_1(x,t)
= (1- \exp(-x/\epsilon_i) )(1-  e^{-x/\epsilon_i} t ) \, .\label{cApr1}
\eea
The interface between the initial fluid and solvent  moves with velocity $u=1$ (i.e. $u^*=u_i^*$). Hence the above solution holds over the region $x \le t/\Da$ ($x^* \le u_i^* t^*)$. The solvent reaches the end of the column when $t_1^* = L^*/u_i^*$ or equivalently $t_1 = {\cal L}^* L/(u_i^*  \tau^*) = L \Da$.

\paragraph{Stage 2}
After $t_1$ equations \eqref{RApr1}-\eqref{cApr1} are valid in the whole column provided there is material to extract in the fibers, that is  while $R \ge R_c$ everywhere. Since the concentration is lowest at the inlet the radius decreases most rapidly there, consequently we may state that this system is  valid provided $R(0,t) \ge R_c$. According to \eqref{RApr1}, to first order,
\bea
\label{Rx0}
R(0,t) \approx   1- t \, ,
\eea
which indicates that the solution is valid for  $t\le t_2= \delta = 1-R_c$ (at least to $\ord{\delta^2}$). We assume that this occurs after the the solvent has first reached the outlet (i.e. $t_2>t_1$, see section \S\ref{LanSec}). We note that $t_2=1-R_c$ corresponds to $\bar{t}=1$ which indicates that the time-scale, the time for the radius to reduce from the initial to the final value, is well chosen.  

One final restriction is given by the fact that the concentration must be below the saturation value, here scaled to unity. In view of \eqref{cApr1} this is satisfied for all times. This indicates that even on this slow time-scale the extraction rate is much slower than the flow, so the solvent passes through the column without ever reaching its saturation value.

\paragraph{Stage 3}

For $t>1-R_c$  extraction occurs beyond some point $x=s(t) > 0$. This is defined as the final point where $R=R_c$ ($\bar{R}=1$), for $x > s$ the fibres are still coated, $R >R_c$. The position $x=s(t)$ will gradually move along the column as material is removed. Consequently, in terms of the new variable $\bar{t}$, for $\bar{t}>1$, we must deal with a moving boundary problem such that
for $x \le s(\bar{t})$
\bea
c(x,\bar{t})  = 0\, ,\quad \bar{R}(x,\bar{t})=1 \, .
\eea
While
for $x \ge s(\bar{t})$ equations
(\ref{ReqnPerB}-\ref{EpsEqnPer}) determine the variables.
At the moving interface continuity of flux requires $c(s(\bar{t}),\bar{t})=0+\ord{\Pe^{-1}}$ and continuity of the radius $\bar{R}(s(\bar{t}),\bar{t})=1$.

To deal with the moving boundary   we switch to a coordinate
system moving with the boundary, $\eta = x-s(\bar{t})$,  and seek a travelling wave form. In terms of the new co-ordinate we define concentration and radius functions $c(x,\bar{t})=f(\eta)$, $\bar{R}(x,\bar{t})=g(\eta)$, where $f(0)=0$, $g(0) = 1$. Sufficiently far ahead of the boundary $\eta \ra \infty$ the solvent is saturated and no material is being extracted, $f=1$, $g=0$. (Note, this is a theoretical point, complete saturation occurs far past the end of the column as $x \ra \infty$ but it is required for the mathematical solution. This does not affect the validity of the result, but  is required to ascertain the velocity.)

Expanding $f(\eta)\sim f_0(\eta)+\delta f_1(\eta)+\cdots$, $g(\eta) \sim g_1(\eta)+\ldots$ and substituting in \eqref{ReqnPerB}-\eqref{EpsEqnPer} the leading and first order terms satisfy
\bea
\epsilon_i \pad{f_0}{\eta} =  1-f_0 \, , \qquad
- \nd{s}{\bar{t}} \pad{g_1 }{\eta}    = 1-f_0 =  \epsilon_i \pad{f_0}{\eta}  \, , \qquad \epsilon_i \pad{f_1}{\eta} =
- g_1(1-f_0)- f_1
\, ,
\eea
where $f_0(0)=f_1(0)= 0$, $g_1(0)=1$. A strong restriction to travelling wave forms is that the speed  $\textrm{d}s/\textrm{d}\bar{t} = v$ is constant, this may only be verified once a solution is obtained. Integrating and applying the boundary conditions gives
\bea
f_0 &=& 1- e^{-\eta/\epsilon_i} \, , 
\\
f_1 &=& \left[\frac{\epsilon_i}{v}\left(  e^{-\eta/\epsilon_i} -1 \right) - \frac{v-\epsilon_i}{\epsilon_i v} \eta \right] e^{-\eta/\epsilon_i}
\, , \qquad
g_1 = 1- \frac{\epsilon_i}{v}\left( 1- e^{-\eta/\epsilon_i} \right) \, .
\eea
In the far-field, $\eta \ra \infty$, the fluid is saturated and we see that $f \ra 1$ is automatically satisfied by the above solution. The condition on the radius $\bar{R} \ra 0$ requires $g_1 \ra 0$. This determines the velocity $v = \epsilon_i$ and so
\bea
f_1 &=& \left(  e^{-\eta/\epsilon_i}   - 1\right) e^{-\eta/\epsilon_i}
\, , \qquad
g_1 =   e^{-\eta/\epsilon_i}  \, .
\label{f1g1t2}
\eea
The fact that we have found a consistent solution indicates that the supposition of constant speed $v= \epsilon_i$ was correct. Then, since $\dd s/\dd \bar{t} = v$, we may now write $s(\bar{t}) = \epsilon_i  (\bar{t}-1)$, after applying $s(1)=0$.
To summarize, the solution in this third stage in terms of $t$ is given by
\begin{equation}
\label{cRStage3App}
\begin{split}
c(x,t) &\sim \left(1- e^{-x/\epsilon_i} e^{({t}/\delta- 1)}\right) \left(1-  \delta  e^{-x/\epsilon_i} e^{({t}/\delta- 1)} \right)
\, , \\
R(x,t) &\sim   1-\delta  e^{-x/\epsilon_i} e^{({t}/\delta- 1)} \, ,
\end{split}
\end{equation}
provided
$$x\ge s(t)=\frac{\epsilon_i}{1-R_c}(t -1+R_c),$$
while for $x<s(t)$ it is simply given by $R(x,t)=R_c$ and $c(x,t)=0$.

This solution holds until  the wave reaches the end of the column, $x=L$. Hence, we define the end of the process as the time $t_f$ in which the wave reaches the outlet, that is $s(t_f)=L$ that is given by
$$ 
t_f=(1-R_c)\left(\frac{1}{\epsilon_i}+1\right)\, .
$$

\subsection{Model for two distinct solubilities}\label{aprox2}
We now consider the case where the material to be extracted has two distinct solubilities with the switch occurring when $R=R_w$ (i.e. $\bar{R}=(1-R_w)/(1-R_c)$). The reduced forms of \eqref{ReqnPerB} and \eqref{ceqnPerB} may be written 
\bea
\pad{\bar{R}}{\bar{t}} & =& \chi-c\,\, \,  ,\qquad \qquad \epsilon_i \pad{c}{x} =  (1-\delta\bar{R})(\chi-c)\,\, \, ,
\eea
where
\bea
\chi = \left\{\begin{array}{cc}
c_w & \textrm{if $\bar{R}\geq (1-R_w)/(1-R_c)$}\\
    1 & \textrm{if $\bar{R}<(1-R_w)/(1-R_c)$}
\end{array}\right.\, .
\eea
\paragraph{Stages 1 and 2}
Until the radius at $x=0$ reaches the switching value, $R_w$, the process involves a single solubility material and so is identical to that studied in \ref{AppA1}. Hence the radius and concentration are defined by (\ref{RApr1}, \ref{cApr1}) and the only difference is that Stage 2 ends when $R(0,t_2)=R_w$,  ($\bar{R}(0,\bar{t)}=(1-R_w)/(1-R_c)$). From equation \eqref{RApr1} we find
\bea
\label{t2neweqAp}
t_2= 1-R_w \, .
\eea

\paragraph{Stage 3}
For $t > t_2$ extraction occurs with two different solubilities. We define the interface by $x=s_1(t)$ such that $R(s_1(t),t)=R_w$ and for $x<s_1(t), \chi=c_w$ while for $x>s_1(t), \chi=1$.

Again we seek a travelling wave form and so introduce the moving coordinate $\eta=x-s_1(\bar{t})$ such that $v_1(\bar{t})=\textrm{d} s_1/\textrm{d}\bar{t}$. As before we seek  solutions of the form $c(x,\bar{t})=f(\eta)$ and $\bar{R}(x,\bar{t})=g(\eta)$ with the conditions that $f(0)=c(s_1(\bar{t}),\bar{t})=A$, where $A$ is an unknown constant, and $g(0)=\bar{R}(s_1(\bar{t}),\bar{t})=(1-R_w)/(1-R_c)$. Expanding again in powers of $\delta$ leads to 
\bea
 \epsilon_i \pad{f_0}{\eta} = \chi-f_0\, , \quad -v_1 \pad{g_1}{\eta} =  \chi-f_0\, , \quad \epsilon_i \pad{f_1}{\eta} + f_1 =-g_1(\chi-f_0)\, ,
\eea
with initial conditions given by $f_0(0)=A$, $f_1(0)=0$  and $g_1(0)=(1-R_w)/(1-R_c)$, and
\bea
\chi = \left\{\begin{array}{cc}
    c_w & \textrm{if $\eta \leq 0$}  \\
    1 & \textrm{if $\eta > 0$}
\end{array}\right.\, .
\eea
The appropriate solutions are 
\bea 
f_0(\eta) &&=\chi + (A-\chi) e^{-\eta/\epsilon_i}\, ,\\
g_1(\eta) &&=\frac{1-R_w}{1-R_c} + \frac{\epsilon_i}{v_1}(A-\chi) (1-e^{-\eta/\epsilon_i})\, ,\\
f_1(\eta) &&=\frac{A-\chi}{\epsilon_i}\left[\left(\frac{\epsilon_i}{v_1}(A-\chi) +\frac{1-R_w}{1-R_c}\right)\eta 
             -\frac{\epsilon_i^2}{v_1}(A-\chi)\left( 1-e^{-\eta/\epsilon_i}\right)\right]e^{-\eta/\epsilon_i}\, .
\eea 
To determine $v_1$ and $A$ we first note that sufficiently far ahead the concentration must approach its saturation value and so there can be no erosion, $c=R=1$. The condition $c=1$ is automatically satisfied as $\eta \ra \infty$ while $R=1$ ($\bar{R}=0$) requires  $g(\eta)\ra 0$ as $\eta\ra\infty$. This gives
\bea 
\label{vel1}
\frac{1-R_w}{1-R_c} + \frac{\epsilon_i}{v_1}(A-1)  = 0\, .
\eea
A second necessary condition may be found by assuming the solution is bounded at either end. As $\eta \ra \infty$ it is already bounded by $c=R=1$. A long way downstream, $\eta \ra -\infty$ the negative exponential would blow up unless $A=\chi = c_w$ (which removes the exponential terms from $f_0, f_1, g_1$) and then from \eqref{vel1} 
\bea 
v_1=
\epsilon_i(1-c_w) \frac{1-R_c}{1-R_w}  \, .
\eea

Substituting for $A, v_1, \chi$ we now obtain
\bea 
f_0(\eta) &&=\left\{\begin{array}{ll}
                c_w & \textrm{if $\eta<0$}\\
                1 - (1-c_w) e^{-\eta/\epsilon_i} & \textrm{if $\eta\geq0$}
            \end{array}\right. \, ,\\
g_1(\eta) &&=\left\{\begin{array}{ll}
                (1-R_w)(1-R_c)^{-1} & \textrm{if $\eta<0$}\\
                (1-R_w)(1-R_c)^{-1}e^{-\eta/\epsilon_i} & \textrm{if $\eta\geq0$}
            \end{array}\right. \, ,\\
f_1(\eta) &&=\left\{\begin{array}{ll}
                0 & \textrm{if $\eta<0$}\\
                -(1-c_w)(1-R_w)(1-R_c)^{-1}\left(1-e^{-\eta/\epsilon_i}\right)e^{-\eta/\epsilon_i} & \textrm{if $\eta\geq0$}
            \end{array}\right. \, .
\eea

Obviously, in reality the column has finite length. The mathematical artifice of imposing infinite boundaries therefore introduces an error which depends on the size of the exponential. If we were to impose the condition $f=0$ at the inlet, $\eta = -s_1$, and consider just the leading order then $f_0(-s_1)=0$ requires $A = c_w (1-e^{-s_1(t)/\epsilon_i})$. This indicates that $A$ in fact varies with time (which is not permitted under the travelling wave assumption) by neglecting this the error in our solution is of the order $e^{-s_1(t)/\epsilon_i}$. So the error due to setting the boundary at negative infinity is greatest  when $s_1$ is close to zero (in fact $s_1=0$ marks the transition from one state to another and we should expect the travelling wave to fail here) but
decreases exponentially with increasing $s_1$. The constant values of the concentration and radius (to this order) for $\eta \le 0$ are also a consequence of the boundary position. The fluid has had plenty of time to reach its saturation value. In general $c_w \ll c_s$ so again the error is small, in \S \ref{LanSec} we see that $c_w \sim 0.1 c_s$ so we may expect maximum errors of the order 10\%. But again this decreases exponentially with $s_1$.

Stage 3 finishes when  the wave reaches the column exit. Noting that
\bea
s_1= v_1 (\bar{t}- \bar{t}_2) = \frac{v_1}{\delta}(t-t_2)
\eea
and applying   $s(t_3)=L$ requires
\bea 
t_3 = (1-R_w) \left(1+ \frac{L}{\epsilon_i(1-c_w)}\right) \, .
\eea
Note, a second possible scenario is that the inlet material is stripped before this wave reaches the outlet. In the case where $c_w \ll c_s$ this is unlikely to happen but in other situations it may be possible. The above solution will not capture this form. A different form of analysis would then be required to deal with the inlet region and match this to the travelling wave. Here, we use the above solution.

\paragraph{Stage 4}
In this stage we seek a travelling wave modelling the removal of the second lanolin layer. This final stage has two distinct components. First, material is removed throughout the column until $R(0,t)=R_c$, subsequently we have a new moving front problem where the core is slowly stripped and the process ends. 

With the current level of approximation, at the start of Stage 4 the radius is $R_w$ everywhere. So during this stage we may proceed as in  Stage 2  for the single solubility model to find the first order solutions
\bea
c(x,t) =  c_w(1-e^{-x/\epsilon_i})\, , \quad R(x,t) = R_w-c_w(t-t_3)e^{-x/\epsilon_i}\, .
\eea
The material at the inlet is  completely eroded when $R(0,t_4)=R_c$, so
\bea
t_4=t_3+\frac{R_w-R_c}{c_w}=(1-R_w) \left(1+\frac{L}{\epsilon_i (1-c_w)}\right)+\frac{R_w-R_c}{c_w}\, .
\eea
\paragraph{Stage 5}
This stage deals with the final stripping process, which has a moving front  defined by $R(s_"( {t}), {t})=R_c$  (with $s_2( {t}_4)=0$). This analysis follows almost exactly that of Stage 3 for the single solubility model, with a slight change in notation.  Introducing the new travelling wave co-ordinate $\zeta=x-s_2(\bar{t})$ (and working in the $\bar{t}$ time-scale) we write  $c(x,\bar{t})=f(\zeta)= f_0(\xi)+\delta f_1(\zeta)+\cdots$ and $\bar{R}(x,\bar{t})=g(\zeta)= g_1(\zeta)+\cdots$. This leads to
\bea
\epsilon_i \pad{f_0}{\zeta}=c_w-f_0 \, ,\quad -v_2 \pad{g_1}{\zeta}=c_w -f_0 \, , \quad \epsilon_i \pad{f_1}{\zeta}=- g_1 (c_w-f_0) -f_1 
\eea
where  $v_2=\textrm{d} s_2/\textrm{d}\bar{t}$. These are subject to $f_0(0)=f_1(0) =0$ (the solvent is clean before the front is reached) and $g_1(0)=1$ (the fibres have reached its core radius, $R_c$). In the far-field, $\zeta \ra \infty$ we have $g_1 \ra (1-R_w)/(1-R_c)$ to determine the velocity. 

Following the analysis of Stage 3 single solubility 
determines
\bea
v_2=\dfrac{1-R_c}{R_w-R_c}c_w\epsilon_i\, ,
\eea
and so
\bea
s_2(t)= \frac{t-t_4}{R_w-R_c}\,c_w\epsilon_i\, .
\eea
To leading order
\bea
c(x,t)&\sim & c_w(1-e^{-(x-s_2(t))/\epsilon_i})\, ,\\
R(x,t)&\sim & R_c+(R_w-R_c)(1-e^{-(x-s_2(t))/\epsilon_i})\,,
\eea
and the process ends  when $s_2(t_f)=L$ such that
\bea
t_f 
&=&(1-R_w)\left(1+\frac{L}{\epsilon_i (1-c_w)}\right)+\frac{R_w-R_c}{c_w}\left(1+\frac{L}{\epsilon_i}\right)\, .
\eea

\newpage

\bibliography{biblioEx}

\begin{thebibliography}{10}
\expandafter\ifx\csname url\endcsname\relax
  \def\url#1{\texttt{#1}}\fi
\expandafter\ifx\csname urlprefix\endcsname\relax\def\urlprefix{URL }\fi
\expandafter\ifx\csname href\endcsname\relax
  \def\href#1#2{#2} \def\path#1{#1}\fi

\bibitem{Huang12}
Z.~Huang, X.-H. Shi, W.-J. Jiang, Theoretical models for supercritical fluid
  extraction, Journal of Chromatography A 1250 (2012) 2--26.

\bibitem{Bakkali2008}
F.~Bakkali, S.~Averbeck, D.~Averbeck, M.~Idaomar, Biological effects of
  essential oils – {A} review, Food and Chemical Toxicology 46~(2) (2008)
  446--475.

\bibitem{Sanda2021}
B.~Sanda, I.~Liliana, Natural dye extraction and dyeing of different fibers: a
  review, John Wiley \& Sons, Ltd, 2021, Ch.~4, pp. 113--135.

\bibitem{Zhang2018}
Q.-W. Zhang, L.-G. Lin, W.-C. Ye, Techniques for extraction and isolation of
  natural products: A comprehensive review, Chinese Medicine 13~(1) (2018)
  1--26.

\bibitem{Thewlis1977}
J.~Thewlis, Lanolin for cosmetic applications, Agro Food Industry Hi-Tech
  May/June (1977) 14--20.

\bibitem{WikiLanolin}
Lanolin, \url{https://en.wikipedia.org/wiki/Lanolin#Applications}, last
  accessed: 04-10-2021.

\bibitem{Sovova94}
H.~Sovov\'a, Rate of the vegetable oil extraction with supercritical carbon
  dioxide - i. modelling of extraction curves, Chemical Engineering Science
  49~(3) (1994) 409--414.

\bibitem{McHugh13}
M.~McHugh, V.~Krukonis, H.~Brenner, Supercritical Fluid Extraction: Principles
  and Practice, Butterworth-Heinemann series in chemical engineering, Elsevier
  Science, 2013.

\bibitem{DeSimone03}
J.~DeSimone, W.~Tumas, Green Chemistry Using Liquid and Supercritical Carbon
  Dioxide, Green Chemistry, Oxford University Press, 2003.

\bibitem{Cheng11}
C.-H. Cheng, T.-B. Du, H.-C. Pi, S.-M. Jang, Y.-H. Lin, H.-T. Lee, Comparative
  study of lipid extraction from microalgae by organic solvent and
  supercritical {CO}$_2$, Bioresource Technology 102~(21) (2011) 10151--10153.

\bibitem{Rajaei05}
A.~Rajaei, M.~Barzegar, Y.~Yamini, Supercritical fluid extraction of tea seed
  oil and its comparison with solvent extraction, European Food Research and
  Technology 220~(3) (2005) 401--405.

\bibitem{Xu00}
Z.~Xu, J.~Godber, Comparison of supercritical fluid and solvent extraction
  methods in extracting $\gamma$-oryzanol from rice bran, Journal of the
  American Oil Chemists' Society 77~(5) (2000) 547--551.

\bibitem{Veress94}
T.~Veress, Sample preparation by supercritical fluid extraction for
  quantification a model based on the diffusion-layer theory for determination
  of extraction time, Journal of Chromatography A 668~(2) (1994) 285--291.

\bibitem{Reverchon99}
E.~Reverchon, J.~Daghero, C.~Marrone, M.~Mattea, M.~Poletto, Supercritical
  fractional extraction of fennel seed oil and essential oil: experiments and
  mathematical modeling, Industrial \& Engineering Chemistry Research 38~(8)
  (1999) 3069--3075.

\bibitem{Perrut97}
M.~Perrut, J.~Clavier, M.~Poletto, E.~Reverchon, Mathematical modeling of
  sunflower seed extraction by supercritical {CO}$_2$, Industrial \&
  engineering chemistry research 36~(2) (1997) 430--435.

\bibitem{Sovova05}
H.~Sovov{\'a}, Mathematical model for supercritical fluid extraction of natural
  products and extraction curve evaluation, The Journal of Supercritical Fluids
  33~(1) (2005) 35--52.

\bibitem{Roy96}
B.~Roy, M.~Goto, T.~Hirose, Extraction of ginger oil with supercritical carbon
  dioxide: experiments and modeling, Industrial \& Engineering Chemistry
  Research 35~(2) (1996) 607--612.

\bibitem{ValverdeReca20}
A.~Valverde, J.~Alvarez-Florez, F.~Recasens, Mathematical modelling of
  supercritical fluid extraction of liquid lanolin from raw wool. {S}olubility
  and mass transfer rate parameters, Chemical Engineering Research and Design
  164 (2020) 352--360.

\bibitem{Valv19}
A.~Valverde, F.~Recasens, Extraction of solid lanoline from raw wool with
  near-critical ethanol modified {CO}$_2$ -- {A} mass transfer model, The
  Journal of Supercritical Fluids 145 (2019) 151--161.

\bibitem{Park75}
J.~Park, O.~Levenspiel, The crackling core model for the reaction of solid
  particles, Chemical Engineering Science 30~(10) (1975) 1207--1214.

\bibitem{Goto96}
M.~Goto, B.~Roy, T.~Hirose, Shrinking-core leaching model for
  supercritical-fluid extraction, The Journal of Supercritical Fluids 9~(2)
  (1996) 128--133.

\bibitem{Levenspiel99}
O.~Levenspiel, Chemical Reaction Engineering, 3rd Edition, John Wiley \& Sons,
  Inc., 1999.

\bibitem{Fior09}
L.~Fiori, D.~Basso, P.~Costa, Supercritical extraction kinetics of seed oil: A
  new model bridging the ‘broken and intact cells’ and the
  ‘shrinking-core’ models, The Journal of Supercritical Fluids 48 (2009)
  131--138.

\bibitem{Rai14}
A.~Rai, K.~D. Punase, B.~Mohanty, R.~Bhargava, Evaluation of models for
  supercritical fluid extraction, International Journal of Heat and Mass
  Transfer 72 (2014) 274--287.

\bibitem{Patel19}
H.~Patel, Fixed bed column adsorption study: a comprehensive review, Applied
  Water Science 9~(45) (2019).

\bibitem{Ahmed18}
M.~J. Ahmed, B.~H. Hameed, Removal of emerging pharmaceutical contaminants by
  adsorption in a fixed-bed column, Ecotoxicology and Environmental Safety 149
  (2018) 257--266.

\bibitem{Myer20a}
T.~G. Myers, F.~Fon, M.~G. Hennessy, Mathematical modelling of carbon capture
  in a packed column by adsorption, Applied Energy 278 (2020) 115565.

\bibitem{Myer20b}
T.~G. Myers, F.~Font, Mass transfer from a fluid flowing through a porous
  media, International Journal of Heat and Mass Transfer 163 (2020) 120374.

\bibitem{Eychenne01}
V.~Eychenne, S.~Sáiz, F.~Trabelsi, F.~Recasens, Near-critical solvent
  extraction of wool with modified carbon dioxide - experimental results, The
  Journal of Supercritical Fluids 21 (2001) 23--31.

\bibitem{Simpson02}
W.~S. Simpson, G.~Crawshaw, Wool: Science and Technology, Woodhead Publishing
  Series in Textiles, Elsevier, 2002.

\bibitem{King26}
A.~T. King, The specific gravity of wool and its relation to swelling and
  sorption in water and other liquids, Journal of the Textile Institute
  Transactions 17~(1) (1926) T53--T67.

\bibitem{Fields11}
P.~R. Fields, T.~L. Chester, A.~M. Stalcup, Viscosity estimation in binary and
  ternary supercritical fluid mixtures containing carbon dioxide using a
  supercritical fluid chromatograph, Journal of Liquid Chromatography \&
  Related Technologies 34 (2011) 995--1003.

\bibitem{Abaroudi99}
K.~Abaroudi, F.~Trabelsi, B.~Calloud-Gabriel, F.~Recasens, Mass transport
  enhancement in modified supercritical fluid, Industrial \& Engineering
  Chemistry Research 38 (1999) 3505--3518.

\bibitem{Puiggene97}
J.~Puiggen\'e, M.~A. Larrayoz, F.~Recasens, Free liquid-to-supercritical fluid
  mass transfer in packed beds, Chemical Engineering Science 52~(2) (1997)
  195--212.

\bibitem{Tan88}
C.-S. Tan, S.-K. Liang, D.-C. Liou, Fluid-solid mass transfer in a
  supercritical fluid extractor, Chemical Engineering Journal 38~(1) (1988)
  17--22.

\end{thebibliography}
\end{document}